\documentclass[final,5p,times]{elsarticle}

\usepackage{amssymb}

\usepackage{amsmath,amssymb,amsfonts}
\usepackage{algorithm,algorithmic}
\usepackage{graphicx}
\usepackage{textcomp}

\usepackage{gensymb}
\usepackage{hyperref}
\usepackage[disable]{todonotes} 
\usepackage{multirow}
\usepackage{booktabs}

\usepackage{mathrsfs}
\usepackage{float}
\usepackage{caption}
\usepackage{subcaption}
\usepackage{array}
\usepackage{makecell}
\usepackage{cleveref}
\usepackage{algorithm}
\usepackage{algorithmic}
\let\labelindent\relax
\usepackage{enumitem}
\usepackage{bm} 

\usepackage{tikz}
\usepackage{pgfplots}
\usepackage{pgfplotstable,booktabs}
\usetikzlibrary{pgfplots.groupplots}
\usepgfplotslibrary{dateplot,fillbetween}
\usepackage{pgf}
\usepgfplotslibrary{colorbrewer}

\newtheorem{remark}{Remark}

\DeclareMathOperator{\Han}{\mathfrak{H}}

\DeclareMathOperator*{\argmin}{arg\,min}

\newcommand{\changeA}[1]{\textcolor{black}{#1}}

\newcommand{\changeC}[1]{\textcolor{black}{#1}}
\newcommand{\changeD}[1]{\textcolor{black}{#1}}

\hypersetup{pdfauthor={Name}} 

 \usepackage[nolist]{acronym} 
\journal{Elsevier}

\begin{document}

\begin{frontmatter}



\title{Adaptive Data-Driven Prediction in a Building Control Hierarchy: \\ A Case Study of Demand Response in Switzerland}


\author[label1]{Jicheng Shi\fnref{label2}\corref{cor1}}
\ead{jicheng.shi@epfl.ch}
\author[label1]{Yingzhao Lian\fnref{label2}}
\author[label1]{Christophe Salzmann}
\author[label1]{Colin N. Jones}

\cortext[cor1]{ Corresponding author}
\fntext[label2]{Jicheng Shi and Yingzhao Lian contributed equally to this work.}

\affiliation[label1]{organization={Automatic Control Lab},
            addressline={EPFL}, 
            city={Lausanne},
            postcode={1015}, 
            state={Vaud},
            country={Switzerland}}

\begin{acronym} 
\acro{agc}[AGC]{Automatic Generation Control}
\acro{mpc}[MPC]{Model Predictive Control}
\acro{as}[AS]{Ancillary Service}
\acro{asp}[ASP]{Ancillary Service Provider}
\acro{bms}[BMS]{Building Management System}
\acro{ddp}[DDP]{Data-Driven Prediction}
\acro{chf}[CHF]{Swiss Franc}
\acro{deepc}[DeePC]{Data-enabled Predictive Control}
\acro{dr}[DR]{Demand Response}
\acro{epfl}[EPFL]{École Polytechnique Fédérale de Lausanne}
\acro{ess}[ESS]{Electrical Storage System}
\acro{hp}[HP]{Heat Pump}
\acro{hvac}[HVAC]{Heating, Cooling and Ventilation}
\acro{io}[I/O]{Input/Output}
\acro{lp}[LP]{Linear Programming}
\acro{lti}[LTI]{Linear Time-Invariant}
\acro{mae}[MAE]{Mean Absolute Error}
\acro{pmv}[PMV]{Predicted Mean Vote}
\acro{ppd}[PPD]{Predicted Percentage Dissatisfied}
\acro{qp}[QP]{Quadratic Programming}
\acro{sfc}[SFC]{Secondary Frequency Control}
\acro{soc}[SoC]{State of Charge}
\end{acronym}

\begin{abstract}
By providing various services, such as \ac{dr}, buildings can play a crucial role in the energy market due to their significant energy consumption. However, effectively commissioning buildings for such desired functionalities requires significant expert knowledge and design effort, considering the variations in building dynamics and intended use. 
\changeC{
In this study, we introduce an adaptive \ac{ddp} layer based on Willems' Fundamental Lemma within the building control hierarchy. This layer is integrated into a bi-level \ac{deepc} structure to achieve diverse control objectives. We validated the proposed method through a building-level case study involving participation in a Swiss \ac{dr} program, conducted in a real building test bed, Polydome.
The \ac{ddp} was utilized to develop a hierarchical controller that provides secondary frequency control on the demand side, with each layer designed to meet specific operational goals. 
Extensive testing with operational data from the Polydome demonstrated that the adaptive \ac{ddp} improves prediction accuracy and reduces tuning efforts compared to standard \ac{deepc} methods. A 52-day continuous experiment in the Polydome, using the tuned parameters, showed that the proposed controller achieved a 28.74\% reduction in operating costs compared to a conventional control scheme.}
Our findings emphasize the potential of the proposed method to reduce the commissioning costs of advanced building control strategies and to facilitate the adoption of new techniques in building control.
\end{abstract}




\end{frontmatter}

\section{Introduction}

Buildings account for approximately 40\% of global energy use and approximately 37\% of carbon emissions, with a large proportion of this potentially controllable through \ac{hvac} systems \cite{united2022}. Recently, indoor climate control and \acf{dr} have received considerable attention for their role in optimizing building operations to enhance efficiency and occupant comfort~\cite{park2018comprehensive}.
In addition, by leveraging the large thermal mass and substantial energy consumption of buildings, flexible operation has demonstrated significant potential in addressing supply-demand imbalances in the power grid.
However, traditional control strategies, such as rule-based controllers, often prove inadequate for these complex tasks.

\ac{mpc} offers an effective solution, particularly with the growing adoption of energy management systems in buildings. MPC predicts building dynamics controlled by HVAC systems and solves optimization problems to determine control actions in a receding-horizon manner. Its inherent ability to integrate economic objectives with comfort constraints makes it highly suitable for indoor climate control and \ac{dr} applications~\cite{drgovna2020all,mariano2021review}. The superiority of MPC over classical building controllers has been validated through various experimental validations~\cite{yang2018state,granderson2018field,sturzenegger2015model,banjac2023implementation,gasser2021predictive,fabietti2018multi,razmara2017building,zhang2022model}. For example, \cite{razmara2017building} demonstrated that MPC reduced grid load ramp rates by 62\% and electricity consumption by 26\% during a demand flexibility task. Similarly, \cite{zhang2022model} conducted a year-long study in a small commercial building, revealing a 12\% reduction in annual energy cost reduction and a 34\% decrease in peak demand compared to conventional methods. 
However, achieving optimal MPC performance requires computationally efficient models with accurate predictions. The complexity of component interactions and the diversity among buildings pose significant challenges to the widespread adoption of MPC in building control~\cite{drgovna2020all,killian2016ten}.

The modeling of building dynamics typically follows two major approaches: first-principles methods and data-driven methods. First-principles models are based on fundamental physical laws, such as heat and mass conservation, and require significant human expertise and domain-specific knowledge of building systems~\cite{zhan2021data}.
To reduce the need for expert involvement, various data-driven techniques have been developed, including Gaussian Processes~\cite{maddalena2022experimental}, Random Forests~\cite{smarra2018data}, and Neural Networks~\cite{ferreira2012neural}.
While these methods take advantage of the abundant data collected from \ac{bms}, they often require large datasets and involve complex nonlinear dynamics~\cite{zhan2021data}, which can introduce additional computational challenges during MPC implementation~\cite{maddalena2022experimental}.

In contrast, linear data-driven models are characterized by their simplicity of identification, efficient sampling and ease of computation within the MPC framework.~\cite{privara2011model,di2022lessons,bunning2022physics,stoffelreal}. For example, \cite{privara2011model} reported 17-24\% energy savings using a state-space model identified via subspace methods~\cite{van2012subspace} compared to a baseline. Similarly, \cite{bunning2022physics} compared a physics informed linear model with Random Forests and Input Convex Neural Networks in two bedrooms, finding comparable control performance with reduced training and computational demands. 
Similar results were observed in~\cite{stoffelreal}, which compared Multiple Linear Regression, Gaussian Processes, and Artificial Neural Networks.

Recently, a non-parametric linear data-driven method for MPC, based on Willems' Fundamental Lemma~\cite{willems2005note}, has received great attention in various domains~\cite{huang2019decentralized,shi2022data,wang2022deep,bilgic2022toward} and has also been adapted for building control~\cite{chinde2022data,yin2024data,lian2021adaptive}. 
This approach uses historical \ac{io} data to directly construct what is known as \acf{deepc}, which eliminates the need for traditional modeling steps and reduces commissioning effort~\cite{coulson2019data}. Compared to subspace identification methods~\cite{van2012subspace}, DeePC does not require a state observer and is less sensitive to the choice of system order. 
For example, \cite{chinde2022data} demonstrated that DeePC performed comparably to a linear-model-based MPC in an office building simulation study. \cite{yin2024data} implemented a stochastic DeePC that outperformed a traditional linear-model-based MPC in both comfort (90\%) and energy savings (8\%). Our previous research~\cite{lian2021adaptive} showed that a bi-level DeePC, using only two days of I/O data, effectively controlled indoor climate in a standalone building with a heat pump at EPFL, reducing energy consumption by 18.4\% compared to an industrial control method.


Despite these advances, there are several gaps in the practical commissioning of DeePC in buildings. Firstly, DeePC's control performance is highly dependent on hyperparameters, the tuning of which requires additional human involvement. Due to the strong integration of prediction and control optimization, tuning often requires either closed-loop testing~\cite{chinde2022data,di2022lessons} or prior knowledge~\cite{yin2024data}. Secondly, existing research on DeePC has primarily focused on short-term indoor climate control, with relatively little investigation of more complex, long-term experiments. 
In long-term operations, a fixed linear model may be inadequate for capturing the slowly time-varying nature of buildings, which is influenced by factors such as weather conditions, occupant behavior, and changes in HVAC system efficiency and operational points~\cite{privara2013building}.
One exception is our previous work~\cite{lian2021adaptive}, where DeePC was adaptively updated using the most recent data in a 20-day experiment. However, this study did not address more complex control scenarios, such as DR, nor did it evaluate the quality of new data updates. In fact, \cite{lian2023physically} highlights that adaptive updates may  sometimes lead to predictions that violate physical laws, potentially resulting in an overestimation of building flexibility.

\changeC{
This study addresses the existing gaps by proposing an adaptive \acf{ddp} layer with a straightforward tuning process to generate flexible bi-level DeePC formulations for various building control objectives. The method is empirically validated through a complicated building-level DR case study in a real building equipped with a \ac{hp}. 
Please note that building-level DR has been demonstrated as both practical and valuable in previous research~\cite{fabietti2018multi,gorecki2017experimental}. Indeed, Swissgrid is advancing a customer-centric and decentralized participation approach with its DR platform, Equigy~\cite{equigy}. 
The key contributions of this study are summarized as follows:}
\begin{itemize}
    \item \changeC{We introduce the DDP layer in the bi-level DeePC structure, serving as a versatile interface for various prediction and control objectives. An adaptive update pipeline is proposed for time-varying approximation of building dynamics.}
    \item \changeC{We conducted a building-level DR case study, focusing on the provision of demand-side SFC services in the Swiss \ac{as} market. The DDP and bi-level DeePC were employed to design a hierarchical SFC controller, which can efficiently handle forecast errors and process disturbances through convex robust optimization.}
    \item  \changeC{Operational data from Polydome, a stand-alone building on the EPFL campus, were used to validate the prediction performance of the adaptive DDP. A sensitivity analysis demonstrated that the DDP requires less tuning effort compared to the standard DeePC.}
    \item The case study was conducted in a 52-day continuous experiment in the Polydome, utilizing the \ac{hp} and a simulated ESS. The SFC controller effectively coordinated subsystems, delivering high-quality power tracking performance. It achieved operating cost savings of 24.64\% and 28.74\% compared to two standard industrial building control baselines. To the best of the authors’ knowledge, this study is the first to validate an adaptive linear data-driven predictive control method for building demand response over such an extended period.
\end{itemize}

\changeC{
The structure of this paper is as follows. Section~\ref{sect:II} introduces the DDP layer, detailing its adaptive update scheme and the construction of the bi-level DeePC.
Section~\ref{sect:IV} describes our case study setup, including the SFC service in the Swiss AS market, the Polydome testbed, and the hierarchical SFC controller using the DDP and bi-level DeePC.
In Section~\ref{sect:V}, we present prediction tests and sensitivity analysis of the adaptive DDP, followed by the results of a  52-day experiment conducted in the Polydome.}
These results are then compared with the outcomes from two alternative operational scenarios using conventional control methods. We conclude this paper with a detailed discussion in Section~\ref{sect:VI} and a conclusion in Section~\ref{sect:VII}.

All acronyms and mathematical symbols used in this paper are listed in Table~\ref{tab:notation1} and~\ref{tab:notation2}.

\begin{table}[!ht]
\centering
\footnotesize
\begin{tabular}{  b{1.5cm}  b{6.2cm} } 
 \hline \\ [-1.5ex]
 Acronym & Definition \\ [0.5ex] 
 \hline \\ [-1.5ex]
 AGC & \acl{agc}\\
 AS &  \acl{as} \\
 ASP & \acl{asp}\\ 
 BMS & \acl{bms}\\ 
 CHF & \acl{chf} \\
 DDP & \acl{ddp} \\
 DeePC & \acl{deepc} \\
 DR & \acl{dr} \\
 EPFL &  \acl{epfl}\\ 
 ESS & \acl{ess}\\
 HP &  \acl{hp}\\ 
 HVAC & \acl{hvac} \\
 LP & \acl{lp} \\ 
 LTI & \acl{lti} \\
 MPC & \acl{mpc} \\
 MAE & \acl{mae} \\
 PMV & \acl{pmv} \\ 
 PPD & \acl{ppd} \\ 
 QP & \acl{qp} \\
 SFC & \acl{sfc}  \\
 SoC & \acl{soc} \\ [0.5ex] 
\hline
\end{tabular}
    \caption{A summary of acronyms. } 
    \label{tab:notation1} 
    \vspace{1.5em}
\end{table}


\begin{table}[!ht]
\centering
\footnotesize
\begin{tabular}{  b{1.5cm}  b{6.2cm} } 
\hline \\ [-1.5ex]
 Variable & Definition\\ [0.5ex]
 \hline \\ [-1.5ex]
 $y$ & Indoor temperature of the building \\
 $u$ & Electrical power setpoint of the HVAC system\\
 $w$ & Outdoor temperature and solar radiation\\
 $P^{H}$ & Actual consumed power of the HVAC system\\
 $SoC$ & State of charge of the ESS\\
 $P^E$ & Power of the ESS\\
 $\alpha$ & AGC signal \\ 
  $c_{bid}$ & Bid price from Swissgrid \\
 $\gamma$ &  Power flexibility\\
 $\bar{P}$ & Power baseline in the day-ahead market\\ 
 $P$ & Power baseline after intraday transaction\\
 $P_{int}$ & Intraday power transaction\\
 $\hat{P}_{int}$ & Prediction of the intraday power transaction\\
 $e_{track}$ & Tracking error in SFC \\ [0.5ex]
 \hline \\ [-1.5ex]
 Parameter & Definition\\ [0.5ex]
 \hline
 $T$ & Length of the operational data for Hankel matrix \\
 $n_{init}$ & Number of the initial steps in bi-level DeePC \\
 $N$ & Number of the predictive steps in bi-level DeePC \\
 $L$ & Depth of the Hankel matrix \\
 $n_{init}$ & Number of the initial steps in bi-level DeePC \\
 $L_{PE}$ & Order for the persistently exciting condition\\
 $n_x$ & Order of the building system\\
 $y_{min/max}$ & Lower or upper bound of indoor temperature\\
 $u_{min/max}$& Lower or upper bound of HP's power\\
 $SoC_{min/max}$& Lower or upper bound of ESS's SoC\\ 
 $P_{min/max}^E$& Lower or upper bound of ESS's power\\
 $N_{scen}$ & Number of historical AGC signal scenarios \\  [0.5ex]
 \hline \\ [-1.5ex]
 Notation & Take some variable $v$ as an example  \\ [0.5ex]
 \hline
  $v_t$ & $v$’s value at time t \\
  $v_\textbf{d}$ & Data sequence from historical operational data\\
  $n_v$ & Dimension of the vector $v$ \\
  $\mathbf{v}$ & The bold letter denotes the sequences over time\\
  $\mathbf{v}_{init}$ & $t_{init}$-step measured values before some time $t$ \\
  $\mathbf{v}_{pred}$ & $N$-step future values from some time $t$ \\
  $\mathbf{v}(i)$ & The $i$-th variable in the sequence $\mathbf{v}$\\
  $v^{(j)}$ & The value of $v$ in the $j$-th scenario\\
  $0$ & A zero matrix with a proper size \\ 
  $I$ & An identity matrix with a proper size \\ [0.5ex]
  \hline
\end{tabular}
    \caption{A summary of variables, parameters and notations. } 
    \label{tab:notation2} 
\end{table}


 
\vspace{-1em}

\section{Methodology} \label{sect:II}

\subsection{Overview}
\begin{figure*}[!ht]
    \centering
    \includegraphics[width=0.6\linewidth]{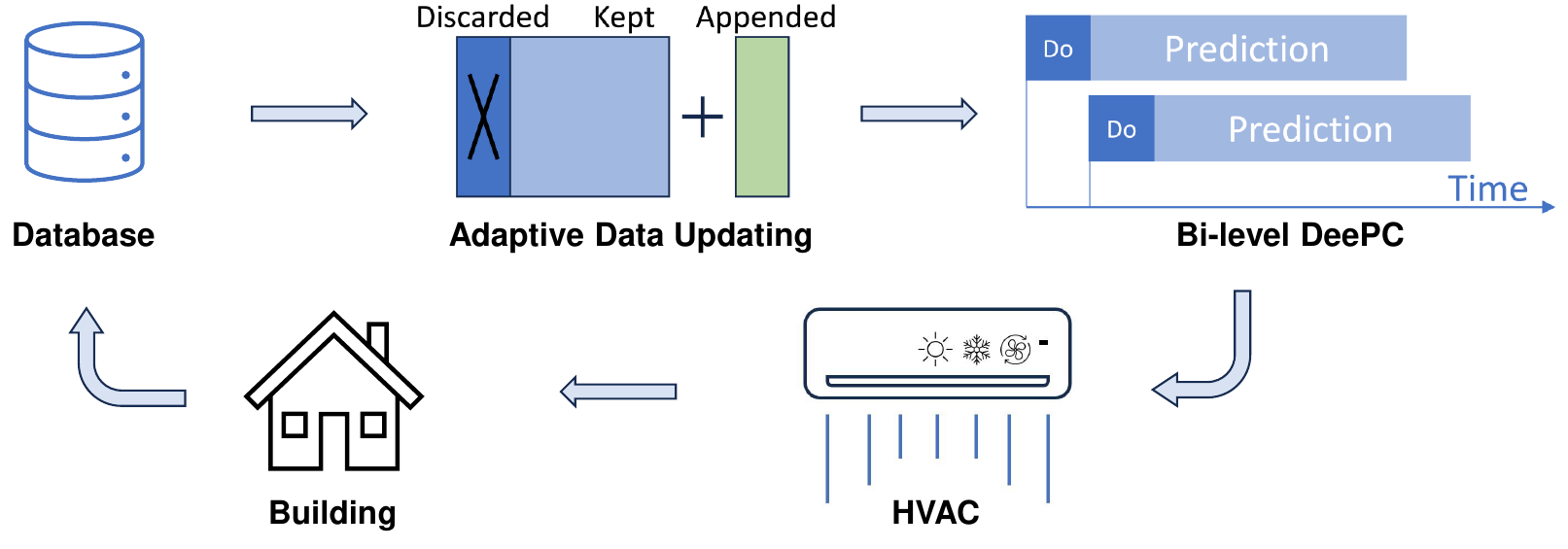}
    \caption{An overview of the core workflow of developing an adaptive predictive controller by the DDP for building systems: an example by the bi-level DeePC}
    \label{fig:core_idea}
\end{figure*}

To give a general idea, the pipeline for implementing a bi-level DeePC controller using adaptive DDP is illustrated in Figure~\ref{fig:core_idea}. This approach utilizes historical operational data collected by the BMS to define the DDP, thereby bypassing  the need for traditional parametric modeling.
Initially, a set of operational data is used to build the first version of bi-level DeePC controller. Subsequently, the controller is updated at each sampling period using the most recent operational data.

\changeD{
In the following subsections, we introduce preliminary knowledge of the adaptive DDP and bi-level DeePC, followed by their formulation and advantages in Sections~\ref{sect:II_LLDP} and~\ref{sect:bilevel}.
Please note that the hierarchical SFC controller used in the DR case study, described in Section~\ref{sect:IV_structure} later, serves as a specific application of the DDP and bi-level DeePC for empirical validation.
}



\subsection{Preliminary knowledge}\label{sect:DDP}

For simplicity, the system outputs, inputs and process noise of building systems are denoted by $y,u,w$, respectively. In the case study, the output $y$ represented the indoor temperature of the building, the process noise $w = [w_1;w_2]$ included the outdoor temperature and solar radiation, and the input $u$ corresponded to the electrical power of the HP. 
Although the unmeasured internal heat gain was not explicitly included in $w$, the adaptive DDP captured the recent building dynamics and demonstrated strong prediction performance, as described in Section~\ref{sect:III_ada}.
Given a $T$-step historical input sequence $u_\textbf{d}:=\{u_i\}_{i=1}^{T}$, its corresponding Hankel matrix of depth $L$, $\Han_L(u_\textbf{d})$, is defined by:
\begingroup\makeatletter\def\f@size{8.5}\check@mathfonts
\begin{align*}
    \Han_L(u_\textbf{d}):=
    \begin{bmatrix}\vspace{-0.1em}
    u_1 & u_2&\dots&u_{T-L+1}\\\vspace{-0.1em}
    u_2 & u_3&\dots&u_{T-L+2}\\\vspace{-0.1em}
    \vdots &\vdots&&\vdots\\\vspace{-0.1em}
    u_{L} & u_{L+1}&\dots&u_T
    \end{bmatrix}\;.
\end{align*}
\endgroup
This Hankel matrixed is further divided into two sub-Hankel matrices:
\begin{align*}
    \Han_L(u_\textbf{d}) = \begin{bmatrix}
        \Han_{L,init}(u_\textbf{d})\\\Han_{L,pred}(u_\textbf{d})
    \end{bmatrix}\;, \label{eqn:Han_sep}
\end{align*}
where $t_{init} + N = L$. The parameters 
$t_{init}$, $N_{}$ and $L$ are user-defined for the DDP, and their selection for the case study is discussed in Section~\ref{sect:V_para}. To maintain consistent notation, the subscript $\cdot_\textbf{d}$ indicates that the data sequence is collected from historical operational data. Similarly, we can define $\Han_L(\cdot)$ for $w$ and $y$.

For a linear time-invariant (LTI) system with an order denoted by $n_x$, the input sequence ${u_\textbf{d}, w_\textbf{d} }$ is termed \textit{persistently exciting} (PE) of order $L_{PE}:= t_{init} + N_{} + n_x $ if the following full row-rank condition is satisfied:
\begin{equation}\label{eqn:PE_check}
    \begin{aligned}
    \text{rank}(\begin{bmatrix}
        \Han_{L_{PE}}(u_\textbf{d}) \\ \Han_{L_{PE}}(w_\textbf{d})
    \end{bmatrix}) =  L_{PE}(n_u+n_w),
    \end{aligned}
\end{equation}
where $n_u$ and $n_w$ denote the dimensions of $u$ and $w$. This condition sufficiently ensures that the input signals in historical data are informative to excite all the representative responses from the target system. Once such an informative input sequence is collected, Willems' Fundamental Lemma can be applied to characterize the I/O responses of the LTI system. Although $n_x$ is generally unknown to us, it can be estimated from historical data~\cite{van2012subspace}, and a conservative overestimate of $n_x$ can be used~\cite{coulson2019data}. For a more detailed discussion about Willems' fundamental lemma, interested readers are referred to~\cite{willems2005note,coulson2019data}.

\vspace{-1em}
\subsection{DDP formulation and its adaptive updates} \label{sect:II_LLDP}
With a historical sequence ${ u_\textbf{d}, w_\textbf{d} }$ that satisfies the PE condition, the DDP for building systems is formulated as a convex \ac{qp} problem, as detailed in~\cite{lian2021adaptive}:
\begin{subequations}\label{eqn:deepc_pred}
\begin{align}
        &\;\mathbf{y}_{pred} = \Han_{L,pred}(y_\textbf{d})g\;,\label{eqn:deepc_pred_sub1}\\
        &g = \argmin_{g_l,\sigma_l} \frac{1}{2}\lVert\sigma_l\rVert^2+\frac{1}{2}g_l^\top\mathcal{E}_g g_l\label{eqn:deepc_pred_sub2}\\
        &\quad\quad\quad\text{s.t.}\;\begin{bmatrix}
        \Han_{L,init}(y_\textbf{d})\\\Han_{L,init}(u_\textbf{d})\\
        \Han_{L,init}(w_\textbf{d})\\
        \Han_{L,pred}(u_\textbf{d})\\
        \Han_{L,pred}(w_\textbf{d})
        \end{bmatrix}g_l=\begin{bmatrix}
        \mathbf{y}_{init}+\sigma_l\\\mathbf{u}_{init}\\ \mathbf{w}_{init}\\\mathbf{u}_{pred}\\\mathbf{w}_{pred}
    \end{bmatrix} \nonumber \;,
\end{align}
\end{subequations}
where the decision variables are $g \in \mathbb{R}^{T-L+1}, \sigma_l \in \mathbb{R}^{t_{init}n_y}$. 
\changeD{The regularization parameter $\mathcal{E}_g$ is user-defined and plays a crucial role in mitigating the negative effects of measurement noise and system non-linearity on prediction accuracy~\cite{lian2021adaptive,dorfler2022bridging,yin2021maximum}.}
Given the previous $n_{init}$-step measurements $\mathbf{u}_{init},\;\mathbf{y}_{init}$ and $\mathbf{w}_{init}$, the problem~\eqref{eqn:deepc_pred} predicts the output sequence $\mathbf{y}_{pred}$ for a specified control input $\mathbf{u}_{pred}$ while accounting for predictive noise $\mathbf{w}_{pred}$.

The solution to the DDP~\eqref{eqn:deepc_pred} can be equivalently derived from the KKT condition, with $\kappa$ as the dual variable for~\eqref{eqn:deepc_pred_sub2}:
\begin{subequations} \label{eqn:deepc_pred2}
\begin{align} 
        &\;\mathbf{y}_{pred} = \Han_{L,pred}(y_\textbf{d})g\;,\label{eqn:deepc_pred2_sub1}\\
        & \;\begin{bmatrix}
        g \\\kappa
        \end{bmatrix} = G^{-1}\begin{bmatrix}\Han_{L,init}(y_\textbf{d})^\top \mathbf{y}_{init}\\\mathbf{u}_{init}\\\mathbf{w}_{init}\\\mathbf{u}_{pred}\\\mathbf{w}_{pred}\end{bmatrix}\;\label{eqn:deepc_pred2_sub2},
\end{align}
\end{subequations}
where 
\begin{subequations}
\begin{align}
    H&:=\begin{bmatrix}
    \Han_{L,init}(u_\textbf{d})\\\Han_{L,init}(w_\textbf{d})\\\Han_{L,pred}(u_\textbf{d})\\\Han_{L,pred}(w_\textbf{d})
    \end{bmatrix} \nonumber \\
    G &:= \begin{bmatrix}
        \Han_{L,init}^\top(y_\textbf{d})\Han_{L,init}(y_\textbf{d})+\mathcal{E}_g &H^\top\\
        H&\textbf{0}
        \end{bmatrix}\;. \nonumber 
\end{align}
\end{subequations}
This reformulation allows us to integrate~\eqref{eqn:deepc_pred} into other optimization control problems without the need to directly solve a bi-level optimization problem. Notably, this equivalent reformulation reveals that the DDP functions as a linear predictor.


Since the DDP is defined directly by measured data, it is logical to update the Hankel matrices with the latest operational data to account for the time-varying nature of building dynamics~\cite{privara2013building}.
During the update process, constructing the Hankel matrix may involve data from multiple time series rather than a single continuous sequence~\cite{van2020willems}. When this occurs, the Hankel matrix is assembled by stacking sub-Hankel matrices from each individual time series~\cite{van2020willems}. This situation often occurs in real-world applications, where control units, such as the \ac{hvac} actuators, may occasionally malfunction or switch operational modes. In our study, the HP switched between the heating and cooling modes, necessitating the use of data from different time series. 
To accurately represent the latest building dynamics, we adjusted the operational dataset to ensure that the Hankel matrix maintained consistent dimensions.

However, directly implementing this adaptive scheme may degrade prediction performance~\cite{lian2023physically}. In~\cite{lian2021adaptive}, additional excitation inputs were added in the closed-loop to mitigate noise-related degradation, but this approach led to increased energy consumption. In contrast, this study employed the following validation tests of data quality to ensure reliable prediction without compromising efficiency:

\noindent \textbf{(1) Persistent Excitation Evaluation:} The updated Hankel matrices can only be accepted if the PE condition~\eqref{eqn:PE_check} is satisfied.  This condition is crucial for the validity of Willems' Fundamental Lemma.


\noindent \textbf{(2) Physical Consistency Test:}
The resulting DDP~\eqref{eqn:deepc_pred2} should align with fundamental physics laws. 
By reformulating the DDP~\eqref{eqn:deepc_pred2} into a linear form:
\begin{align*} 
        \mathbf{y}_{pred} & = \Han_{L,pred}(y_\textbf{d}) G^{-1}\begin{bmatrix}\Han_{L,init}(y_\textbf{d})^\top \mathbf{y}_{init}\\\mathbf{u}_{init}\\\mathbf{w}_{init}\\\mathbf{u}_{pred}\\\mathbf{w}_{pred}\end{bmatrix}\;\\
        & := P_1\mathbf{y}_{init} + P_2\mathbf{u}_{init} + P_3\mathbf{u}_{pred} + P_4\mathbf{w}_{pred},
\end{align*}
the coefficients $P_i$ can be checked a-posteriori.
For example, in the case of a HP operating in cooling mode, an increase in electrical power $u$ should logically lead to a decrease in room temperature. 
Because each column of $P_3$ represents the linear effects of each future input on the future $N$-step outputs, the following equation indicates that the average future temperature will decrease if one future electrical power $\mathbf{u}_{pred}(i)$ increases:
\begin{equation}\label{eqn:PC_check_2}
\begin{aligned}
    \sum_{j=1}^{n_y N} P_3(j,i) < 0
\end{aligned}   
\end{equation}
\changeD{However, formulating the DDP using raw data may violate this physical rule due to noise in the data and the nonlinear dynamics in buildings. This can result in an overestimation of building flexibility~\cite{lian2023physically}.
To mitigate this risk, the DDP was updated only when the following relaxed condition was satisfied in this study:
\begin{equation}\label{eqn:PC_check}
    \begin{aligned}
    \sum_{i=1}^{n_u N} \Gamma(i) & \geq \eta n_u N \\ \text{where } & \Gamma(i) = \left\{
      \begin{array}{ll}
    1, & \text{if} \; \eqref{eqn:PC_check_2} \; \text{is satisfied} \\
        0, & \text{otherwise}
    \end{array}
    \right. ,
    \end{aligned}
\end{equation}
with a selected threshold $\eta \in [0,1]$.
It requires that at least $\eta n_u N$ inputs in the DDP satisfy the physical rule~\eqref{eqn:PC_check_2}. In the case study, we set $\eta=0.8$.
}

\subsection{Build bi-level DeePC by DDP}\label{sect:bilevel}
Incorporating the DDP as a predictive component, the structure of the bi-level DeePC is defined as follows:
\begin{subequations}\label{eqn:deepc_og}
\begin{align}
    \min\limits_{\substack{\mathbf{y}_{pred}\\\mathbf{u}_{pred}}} \;&\; J(\mathbf{y}_{pred},\mathbf{u}_{pred}) \label{eqn:deepc_og_cost}\\
        \text{s.t.} 
        &\; \mathbf{u}_{pred} \in \mathcal{U} \;,\label{eqn:deepc_og_ucons}\\
        &\;\mathbf{y}_{pred} \in \mathcal{Y}\;,\label{eqn:deepc_og_ycons}\\
        & \eqref{eqn:deepc_pred2_sub1},\eqref{eqn:deepc_pred2_sub2} \nonumber\;.
\end{align}
\end{subequations}
Here, $J(\mathbf{y}_{pred},\mathbf{u}_{pred})$ represents the user-defined control objective, while $\mathcal{U}$ and $\mathcal{Y}$ denote  the input and output constraints. A template code for implementing bi-level DeePC for indoor climate control can be found on GitHub\footnote{\href{https://github.com/YingZhaoleo/RISK_src_yingzhao/tree/main/deepc/robust_deepc_bilevel}{https://github.com/YingZhaoleo/RISK\_src\_yingzhao/tree/\\main/deepc}}.

\changeD{
The bi-level structure consists of an upper level optimization~\eqref{eqn:deepc_og} and a lower level DDP prediction problem. unlike the standard DeePC setup~\cite{coulson2019data,chinde2022data}, the DDP operates independently of the upper level optimization. This independence allows for offline tuning using historical data, with the tuned hyperparameters applicable to any upper-level optimization configuration.
In Section~\ref{sect:III_tune}, we conducted prediction tests by real-life operational data, showing that the bi-level DeePC requires less tuning effort than the standard DeePC~\cite{coulson2019data,chinde2022data}.  
Additionally, during the control process, the DDP can be adaptively updated to capture time-varying building dynamics, as exemplified in Algorithm~\ref{alg:deepc_ada}, where the update cycle can be customized for specific applications.
}

\changeD{
The bi-level structure also allows the user to define their preferred choices of $J(\mathbf{y}_{pred},\mathbf{u}_{pred})$, $\mathcal{U}$, and $\mathcal{Y}$.
For example, in~\cite{lian2021adaptive}, a specific bi-level DeePC was implemented for basic indoor climate control. $J(\mathbf{y}_{pred},\mathbf{u}_{pred})$ was set to minimize the total HP power consumption, $\mathcal{U} = \left[ u_{min}, u_{max} \right]$ defined the HP power limits, and $\mathcal{Y}=\left[ y_{min}, y_{max} \right]$ established a comfortable range for indoor temperature. 
In this work, we utilized the DDP and bi-level DeePC to develop the SFC controller for the complex DR case study, as detailed in Section~\ref{sect:IV}.
}

\begin{algorithm} 
  \caption{Adaptive bi-level DeePC}
  \label{alg:deepc_ada} 
{
\begin{itemize} [nolistsep,leftmargin=1.5em]
    \item[1)] (This step is only performed when conducting an update). Retrieve the recent operational data $\{ u_\textbf{d}, w_\textbf{d}, y_\textbf{d}\}$. If the conditions~\eqref{eqn:PE_check} and~\eqref{eqn:PC_check} are satisfied, update the Hankel matrix. 
    \item[2)] \changeA{ Retrieve the most recent $t_{init}$-step measurements $\mathbf{u}_{init}, \mathbf{w}_{init}, \mathbf{y}_{init}$. Solve~\eqref{eqn:deepc_og} and obtain the solution $\mathbf{u}_{pred}^{\ast}$. Apply the first control action as $u_t=\mathbf{u}_{pred}^{\ast}(1)$. }
    \item[3)] Pause until the subsequent sampling time, then return to step 1.
\end{itemize}
}
\end{algorithm}

\section{Case study: setup of DR program, building test bed and SFC controller} \label{sect:IV}

\changeC{
To empirically validate the efficacy of DDP in complex building control tasks, we conducted a building-level DR case study. This section details the setup of the DR case study. Section~\ref{sect:IV_SFC} provides an overview of the target DR program, specifically focusing on SFC in the Switzerland AS market. Section~\ref{sect:IV_testbed} describes the experimental test bed, Polydome, and Section~\ref{sect:IV_structure} explains the construction of the hierarchical SFC controller based on the DDP and bi-level DeePC.
}


\subsection{DR program setup} \label{sect:IV_SFC}

To ensure the reliable and secure operation of the Swiss transmission grid, Swissgrid has established an AS market that includes services such as frequency control and voltage support. 
\changeC{
In this market, the SFC service requires participants, either on the production or demand side, to respond within seconds and adjust their production or consumption rates within 15 minutes to mitigate grid imbalances. In this study, we used DDP and bi-level DeePC to coordinate the stand-alone Polydome and an ESS to provide the SFC service from the demand side. This building-level case study is both practical and valuable, particularly as Swissgrid is advancing customer-centric and decentralized participation in the AS market~\cite{equigy}. Additionally, pilot research has demonstrated significant potential for this approach~\cite{fabietti2018multi,gorecki2017experimental}.
}


Next, two stages in the Swiss SFC setup are introduced: a day-ahead market and intraday tracking. In the day-ahead market stage, Swissgrid collects the bid of power flexibility $\gamma$ and the power baseline $\bar{P}$ from each \ac{asp}. The power flexibility $\gamma$ represents the maximum power flexibility the ASP can alter from the claimed power baseline $\bar{P}$ during the subsequent day. In return, each ASP receives a reward equivalent to $c_{bid}\gamma$, where $c_{bid}$ is the bid price from Swissgrid.

During the intraday tracking stage, Swissgrid sends a time-dependent normalized signal, known as the \ac{agc} signal $\alpha_t$, to the ASPs at time $t$. Each ASP is then required to adjust its power consumption, deviating from the declared power baseline by $\gamma \alpha_t$. At this stage, the baseline can be re-traded up to 45 minutes before delivery in an intraday market, a measure that significantly enhances the building's bidding flexibility~\cite{qureshi2016economic}. At time $t$, the final power baseline $P_{t+3}$ is set after submitting the intraday power transaction $P_{int,t+3}$, assuming a 15-minute sampling period. Therefore, the required consumed power $P_t$ for intraday tracking is defined as: 
\begin{align*} 
P_t = \bar{P}_{t} + P_{int,t} + \gamma \alpha_t 
\end{align*}



\begin{figure*}[!ht]
    \centering
    \includegraphics[width=0.8\linewidth]{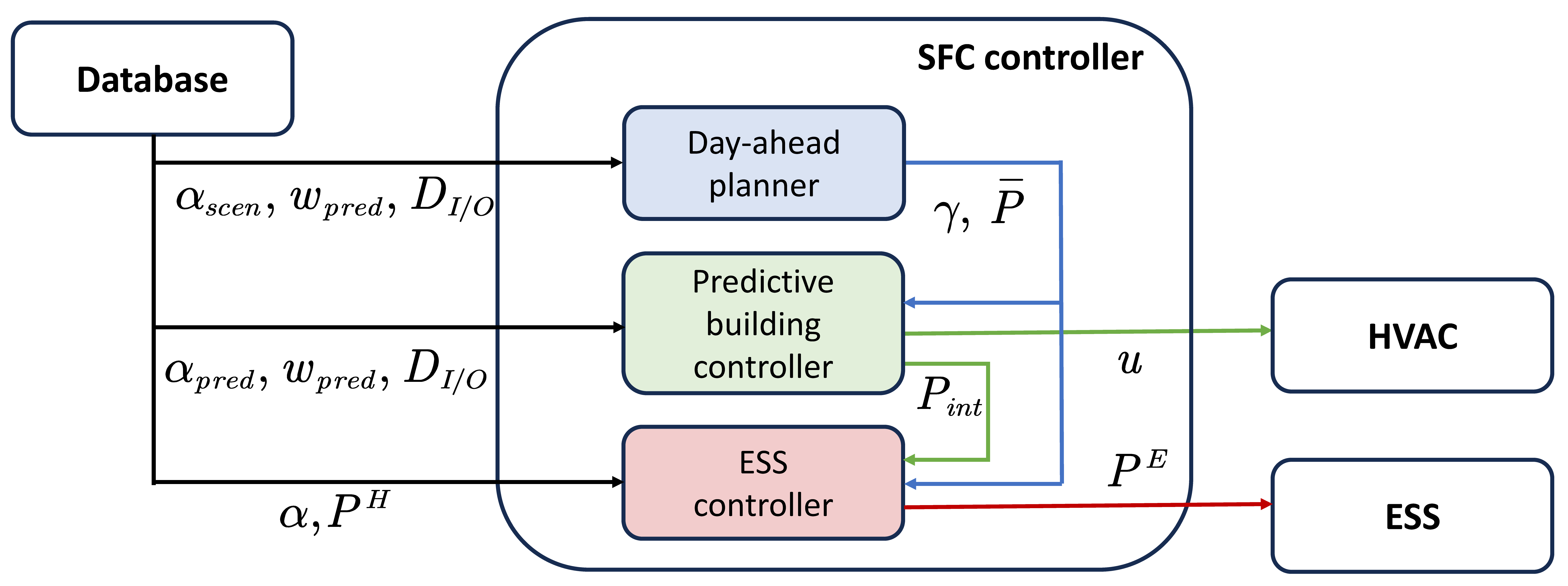}
    \caption{Schematic diagram of the SFC controller: communication inside the controller and communication with external modules}
    \label{fig:3control}
    \vspace{-0.0em}
\end{figure*}

\begin{table*}[!ht]
\begin{center}
\begin{tabular}{  b{2.2cm}  b{1.7cm} b{3.6cm}  b{3.6cm}  b{3.7cm}  }
  \hline
  Layer & Sampling time & Inputs from \newline other layers & Inputs from \newline external modules & Outputs \\ 
  \hline
  Day-ahead Planner  & 1 day & \rule[0.5ex]{3.5ex}{0.5pt}  &  AGC scenarios $\alpha_{acen}$,\newline weather forecast $\mathbf{w}_{pred}$, \newline operational data $D_{I/O}$
  & power flexibility$\gamma$,\newline day-ahead power baseline $\bar{P}$ \\ 
  \hline
  Predictive Building \newline Controller  & 15 minutes & power flexibility$\gamma$,\newline day-ahead power baseline $\bar{P}$
  & AGC prediction $\alpha_{pred}$,\newline weather forecast $\mathbf{w}_{pred}$, \newline operational data $D_{I/O}$
  & HP setpoint $u$,\newline intraday power transaction $P_{int}$ \\  
  \hline
  ESS controller  & 4 seconds & power flexibility$\gamma$,\newline day-ahead power baseline $\bar{P}$, \newline intraday power transaction $P_{int}$  
  & AGC signal $\alpha$,\newline HP power $P^{H}$
  & ESS power $P^{E}$ \\ 
  \hline  
\end{tabular}
\caption{Characteristics of the SFC controller: sampling time, inputs and outputs} \label{tab:3layer}
 \end{center}
 \vspace{-1.5em}
\end{table*}

This study employed a combination of the building and ESS to carry out the SFC task. The building naturally provides a substantial energy capacity to meet the aggregate energy demands of \ac{agc} signals over extended periods, while the ESS can rapidly respond to high-frequency AGC fluctuations. This is critical for achieving high intraday tracking performance. This integrated approach has been shown to yield higher returns at lower costs in the AS market, compared to using only ESS~\cite{fabietti2018multi}. 
In this study, the optimal solution of predicted input $\mathbf{u}_{pred}$ in the bi-level DeePC determined the electrical power setpoint of the HP, and $P^{H}$ denoted the actual power consumed by the HP, as there might be slight deviations from the setpoint.
For the ESS, $SoC$ and $P^E$ denoted the \ac{soc} and power, respectively. Hence, during the intraday tracking stage, the control task aimed to adhere to the required consumed power $P_t$, defined by:
\begin{align*}
P_t = P_{t}^H + P_{t}^E
\end{align*}
If the AGC signal is not perfectly tracked, the ASP must pay a penalty based on the tracking error $e_{track}$. In this study, $e_{track}$ was defined as: 
\begin{align*}
    e_{track,t} = P_t - P_{t}^H - P_{t}^E\;.
\end{align*}

\begin{remark}
This paper uses a passive sign notation, where a positive power baseline indicates consumption, and a positive (negative) AGC signal indicates increased (decreased) power consumption.
\end{remark}

\begin{figure}[!ht]
    \centering
    \includegraphics[width=0.9\linewidth]{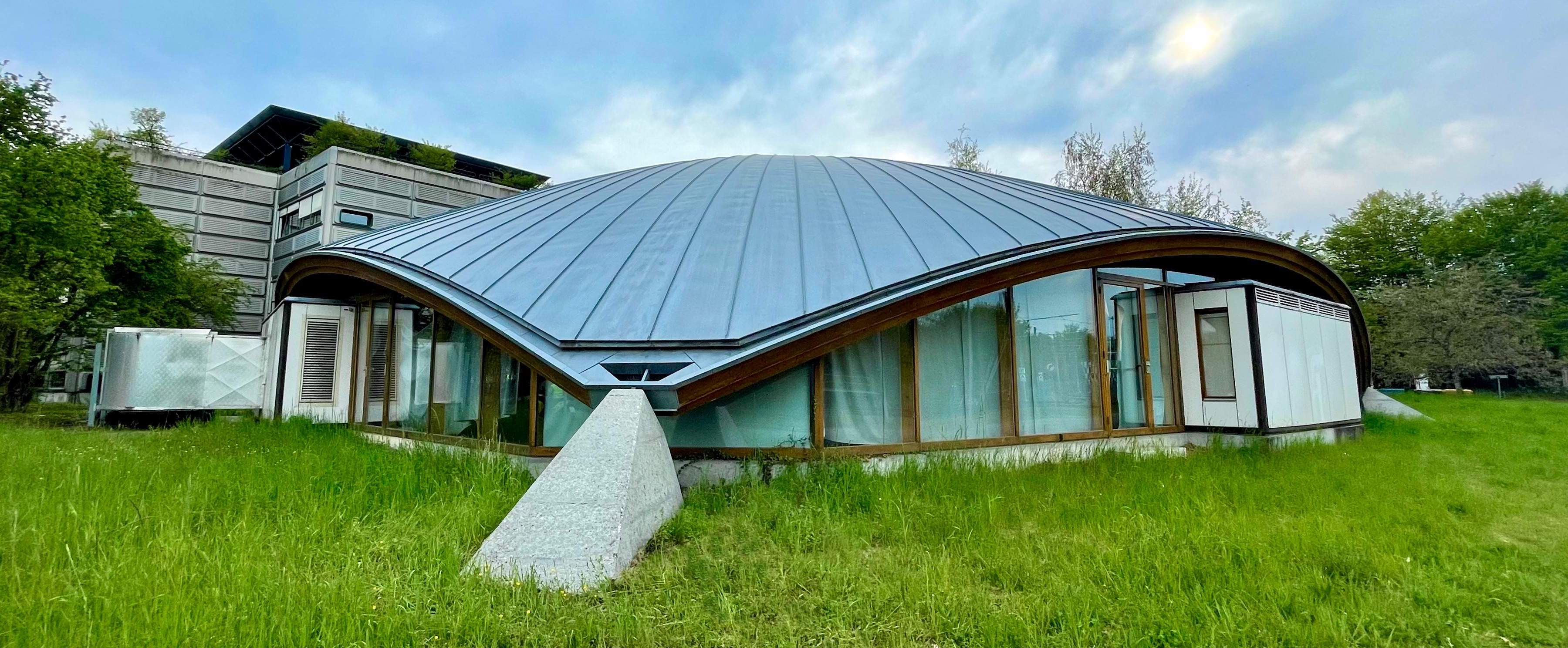}
    \caption{Photo of the Polydome}
    \label{fig:polydome_ext}
\end{figure}

\begin{figure*}[!ht]
    \centering
    \includegraphics[width=0.8\linewidth]{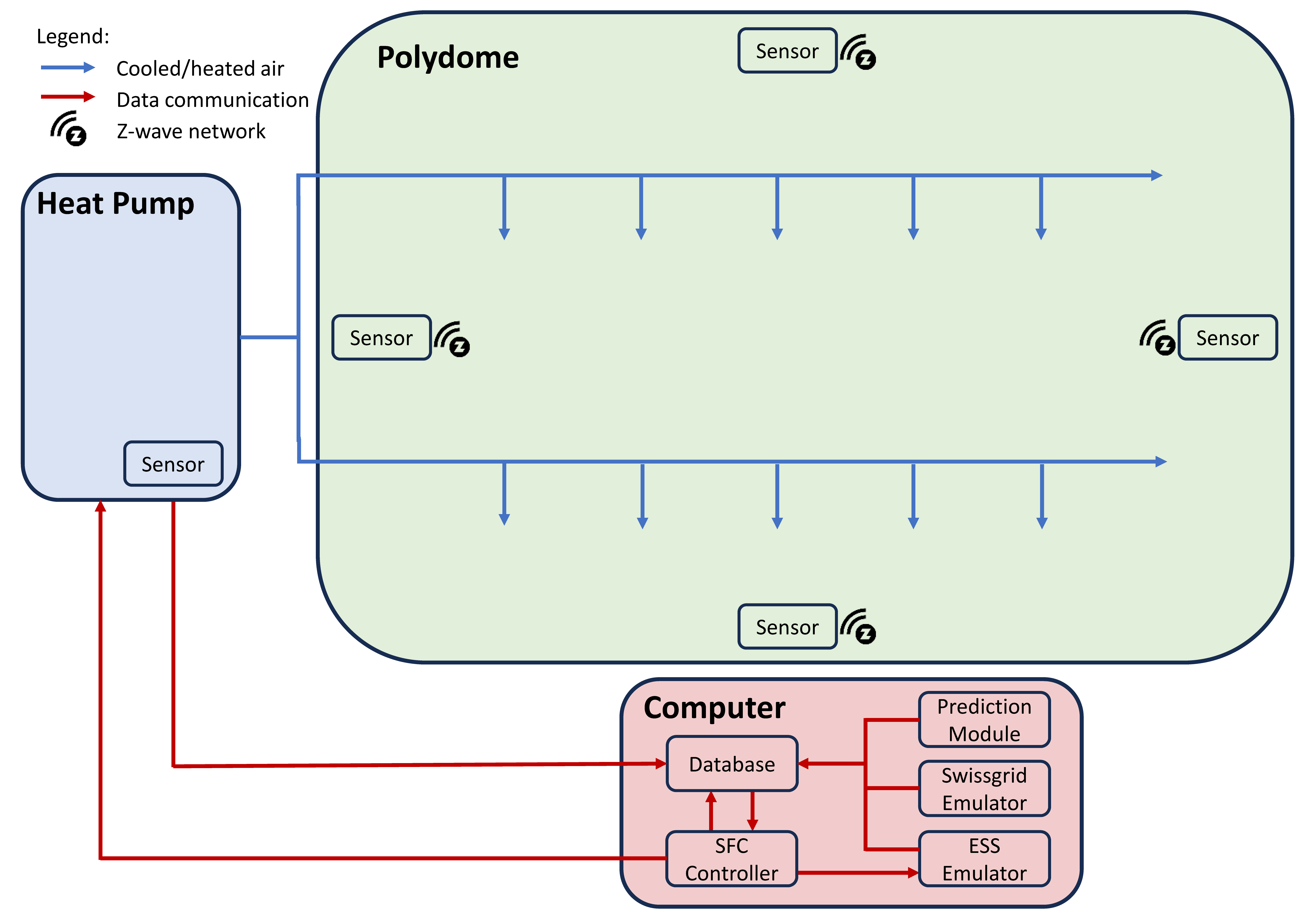}
    \caption{
    The overall testbed architecture, including the main components and data communication directions.}
    \label{fig:testbed}
\end{figure*}

\subsection{Testbed setup} \label{sect:IV_testbed}

This section describes the Polydome, an isolated building of 600 m² used as a lecture, exam and conference room with a capacity of 200 people  (Figure~\ref{fig:polydome_ext}). Figure~\ref{fig:testbed} presents a diagram of the testbed, illustrating the relationships among the major components of the case study. The blue arrow indicates the distribution of cooled/ heated air from the HP to the Polydome, while the red arrow represents the direction of data communication. In the case study, a central computer (iMac with 32 GB 1600 MHz DDR3 memory and a 3.4 GHz Intel Core i7) managed several components. Various wireless sensors were installed in the Polydome, utilizing a Z-wave network for communication with the central computer. The components are detailed as follows.

\textbf{\ac{hp}}: The AERMEC RTY-04 HP operates its ventilation fan continuously, consuming a fixed 2.4 kW, to circulate fresh air throughout the lecture hall via two elongated fabric pipes. The HP can operate in either heating or cooling mode, but only one mode can be active at a time.  Depending on the selected mode, the fan distributes either hot or cool air evenly throughout the Polydome, as illustrated by the blue arrows in Figure~\ref{fig:testbed}.
To ensure stable and automatic operation, the school administration requires a scheduler in the BMS to control the HP mode. The scheduler switches between heating and cooling based on a predefined outdoor temperature threshold and includes a delay of over an hour to prevent frequent mode changes.
\changeD{
It is important to note that we were \textbf{not} authorized to adjust this scheduler or directly alter the mode. While the scheduler could occasionally malfunction, such as maintaining heating at midday in summer, the designed SFC controller was robust enough to handle these situations, as demonstrated in the results analysis in Section~\ref{sect:V_result}.}
Within the BMS, an internal industrial controller uses a bang-bang control rule to stabilize the return air temperature around a predefined setpoint. The HP can be modulated to follow a specified power setpoint $u$, with adjustments allowed every 15 minutes due to compressor constraints. In heating mode, the HP can reach a maximum nominal power of $8.4, kW$ ($7, kW$ in cooling mode), inclusive of the fan's power.

\textbf{Sensors}: The HP's power consumption was monitored using an EMU 3-phase meter~\cite{emu}, which sent the data to the central computer via Ethernet. This monitoring is crucial because the HP's actual power consumption $P_{t}^H$ may differ slightly from the power setpoint $u_t$. For indoor temperature monitoring, four Fibaro Door/Window Sensor v2 units were installed at a height of 1.5 meters on different walls. These sensors transmitted temperature data to the central computer every five minutes through the Z-wave wireless protocol. The building's output, $y$, was calculated as the average temperature value from the four sensors, with their locations shown in in Figure~\ref{fig:testbed}.

\textbf{Database}: A time-series database, InfluxDB 1.3.7~\cite{influxdb}, was used on the central computer to store operational data, including the HP power consumption, indoor temperature, and historical weather data. It also recorded all data related to the SFC control task, such as the AGC signals, power flexibility, and ESS power consumption.

\textbf{SFC Control Module}: The hierarchical SFC controller was implemented on the central computer using MATLAB. This module, detailed in Section~\ref{sect:IV_structure} communicated with the database and ESS emulator, both located on the same computer. It calculated the HP's power setpoint, $u$, and sent the modulated signal to the BMS via the Modbus serial communication protocol~\cite{swales1999open}.

\textbf{Prediction Module}: This module, also hosted on the central computer, provided forecasts for weather conditions and AGC signals, storing the results in the database. It utilized the Tomorrow.io~\cite{tomorrow} API for weather forecasts, including outdoor air temperature and solar radiation data, with predictions extending up to 24 hours ahead and updated every 15 minutes. The API also accessed current weather conditions, which were refreshed every five minutes. Additionally, the module handled AGC prediction tasks, with further details provided in Section~\ref{sect:V_para}.

\textbf{Swissgrid Emulator}: This emulator, hosted on the central computer, randomly selected a sequence of AGC signals from a historical 2020 dataset every midnight. Throughout the day, it sent the AGC signal to the SFC control module every four seconds.
    
\textbf{ESS Emulator}: This component emulated a battery using a discrete-time nonlinear model with a four-second sampling interval, based on the battery model from~\cite{sossan2016achieving}. The simulated battery had an energy capacity of $5kWh$ and a maximum charge/discharge rate of $5kW$. Its dynamics are expressed as:
\begin{align} \label{eqn:ess_dyn}
SoC_{t+1} = SoC_{t} + 0.95(P_{t}^E)_+ - \frac{1}{0.95}(P_{t}^E)_-
\end{align}
where $(P^E)+$ and $(P^E)_-$ denote the power injection and extraction in the battery, respectively. Due to the ESS's ability to precisely and rapidly track the power setpoint, it was assumed that the ESS could inject or extract the required power in sync with the AGC signals, every 4 seconds.



\subsection{SFC control structure} \label{sect:IV_structure}

In this section, we describe the construction of the hierarchical SFC controller based on the DDP and bi-level DeePC, as implemented in the case study. The SFC controller is illustrated in Fig.~\ref{fig:3control}, with its principal attributes summarized in Table~\ref{tab:3layer}. The functionality of each layer is briefly outlined below.

\textbf{Day-ahead planner}: Each day at 23:45, the planner gathered all pertinent information, including operational data $D_{I/O}$, historical AGC scenarios $a_{scen}$, and weather forecasts $\mathbf{w}_{pred}$. Utilizing day-ahead predictions, it formulated an optimization planning problem to determine the power flexibility $\gamma$ and the day-ahead power baseline $\bar{P}$. This problem employed a scenario optimization method that incorporated predictions of intraday power transactions to circumvent overly conservative outcomes.

\textbf{Predictive building controller}: At 00:00 each day, this layer adopted the results from the day-ahead planner and operational data $D_{I/O}$ from the database. Every 15 minutes thereafter, it collected the latest weather forecast $\mathbf{w}_{pred}$ and AGC prediction $\alpha_{pred}$, then solved a worst-case robust predictive control problem to determine the optimal intraday power transaction $P_{int}$ and power setpoint $u$. The modulated signal based on $u$ was then sent to the BMS.

\textbf{ESS controller}: Operating at a 4-second interval this layer collected the current building power consumption $P^H$, the current AGC signal $\alpha$, the power flexibility $\gamma$, and the current baseline $P$. It then calculated the required power $P^E$ and transmitted it to the ESS emulator for AGC tracking.

For convenience, we defined the feasible power sets for the building and the ESS, considering an $N_{}$-step prediction from time $t$. 
The power set of the building based on the DDP was defined as
\begin{align}
    \mathbb{P}^H&(\mathbf{u}_{init},\mathbf{w}_{init},\mathbf{y}_{init},\mathbf{w}_{pred}) \nonumber \\ 
    &= \left\{ \mathbf{u}_{pred}^H \middle|
    \begin{aligned}
        & \eqref{eqn:deepc_pred2} \\
        & u_{min} \leq \mathbf{u}_{pred}(i) \leq u_{max} \\
        &  y_{min} \leq \mathbf{y}_{pred}(i) \leq y_{max}\\
        &\forall i = 1,\dots,N_{}
    \end{aligned}\right\}\;, \label{eqn:SFC_building}
\end{align}
It employed the DDP~\eqref{eqn:deepc_pred2} to predict future temperatures, where $y_{min}$ and $y_{max}$ constrained the indoor temperature to a comfortable range and the box constraint on $\mathbf{u}$ indicated the HVAC's limitations.
For the ESS, it was defined as follows:
\begin{align}
    \mathbb{P}^E&(SoC_{t}) \nonumber \\
    &= \left\{ \mathbf{P}_{pred}^E \middle|
    \begin{aligned}
        & \mathbf{SoC}_{pred}(i) = \mathbf{SoC}_{pred}(i-1) + \mathbf{P}_{pred}^E(i), \\
        & \mathbf{SoC}_{pred}(0) = SoC_{t} \\ 
        & P_{min}^E \leq \mathbf{P}_{pred}^E(i) \leq P_{max}^E, \\
        & SOC_{min} \leq \mathbf{SoC}_{pred}(i) \leq SOC_{max},\\
        &\forall i = 1,\dots,N_{}
    \end{aligned}\right\}\;, \label{eqn:SFC_ESS}
\end{align}
It used a linear approximation of the ESS model~\cite{sossan2016achieving}, with box constraints applied to both the \ac{soc} and the battery power.







\subsubsection{Day-ahead planner}

At 23:45 each day, the day-ahead planner computed the power baseline $\bar{P}$ and the power flexibility $\gamma$ for the next 24 hours, starting at 00:00. 
These values were then transmitted to the other two control layers. In the Swiss AS market, the baseline $\bar{P}$ should be specified for every 15-minute time slot~\cite{fabietti2018multi}, so the prediction step $N$ was set to 96.

The proposed day-ahead planner, built using the DDP and bi-level DeePC, executed the following steps daily at 23:45:
\vspace{1.0em}
\begin{itemize}[topsep=0em,itemsep=0.1em] 
    \item[1.] Retrieve the recent operational data $D_{I/O}$. If the two prerequisites, ~\eqref{eqn:PE_check} and~\eqref{eqn:PC_check}, are met, update the Hankel matrix. 
    \item[2.] Construct the $t_{init}$-step vectors $\mathbf{y}_{init}, \mathbf{u}_{init}, \mathbf{w}_{init}$ from the most recent operational data.
    \item[3.] Retrieve the weather forecast from the API and construct the vector $\mathbf{w}_{pred}$.
    \item[4.] Collect $N_{scen}$ AGC scenarios from historical data, $\bm{\alpha}_{scen}^{(j)}$, $\forall \; j = 1,2,\dots,N_{scen}$. Compute the predicted intraday power transactions $\mathbf{\hat{P}}_{int}^{(j)}$ using~\eqref{eqn:SFC1_agc}, details of which will be explained later.
    \item[5.] Obtain the current \ac{soc} $SoC_{t}$ from the ESS.
    \item[6.] Solve the following scenario optimization problem with a resolution of 15 minutes.    
    \end{itemize}
        \begin{subequations}\label{eqn:SFC1}
        \begingroup
        \allowdisplaybreaks        
        \begin{align}
                &\underset{\substack{\gamma, \mathbf{\bar{P}}, \mathbf{P}^{S,(j)}, \mathbf{u}_{pred}^{(j)}}}{\text{minimize}}
                \;\; &&-\gamma \label{eqn:SFC1_1} \\
                &\text{s.t.} && \nonumber \\
                &\text{(Building system)}  && \mathbf{u}_{pred}^{(j)} \in \mathbb{P}^H \left(\begin{array}{@{}c@{}} \mathbf{u}_{init},\mathbf{w}_{init},\\ \mathbf{y}_{init},\mathbf{w}_{pred}\end{array} \right) \;, \label{eqn:SFC1_3}\\
                &\text{(ESS system)} && \mathbf{P}_{pred}^{E,(j)}   \in \mathbb{P}^E(SoC_{t}) \;,  \\
                & \text{(Total power)} && \mathbf{P}^{(j)} = \mathbf{P}_{pred}^{E,(j)} + \mathbf{u}_{pred}^{(j)} \;, \label{eqn:SFC1_4} \\
                & \text{(Power tracking)}  && \mathbf{P}^{(j)} = \mathbf{\bar{P}} + \mathbf{\hat{P}}_{int}^{(j)} + \gamma \bm{\alpha}_{scen}^{(j)} \;, \label{eqn:SFC1_5} \\
                &\text{(Power flexibility)} && \gamma \geq 0 \;, \label{eqn:SFC1_6} \\
                &\; && \forall \; j = 1,2,\dots,N_{scen} \nonumber
        \end{align}
        \endgroup
        \end{subequations}

\begin{itemize}[topsep=0em]      
    \item[7.] Transmit the optimal $\gamma$ and $\bar{P}$ values to the other two layers.
    \item[8.] Stay idle until the next iteration.
\end{itemize}


\changeD{
The details of the day-ahead planner are discussed as follows, focusing on how it combines the DDP with scenario programming in~\eqref{eqn:SFC1}. Here, $\bm{\alpha}_{scen}^{(j)}$ symbolized the AGC signals from a complete historical day as $j$-th scenario, with related variables sharing the same superscript $^{(j)}$. While the actual AGC signals $\bm{\alpha}$ for the following day were unknown, it was assumed that the AGC signals exhibit a consistent statistical feature, enabling scenario robust optimization. This approach has two key benefits: it produces decisions robust against future realizations of AGC signals by optimizing over numerous past scenarios; it is less conservative than worst-case robust optimization methods~\cite{shapiro2021lectures}, thereby preventing significant reductions in economic returns. 
}

In the objective~\eqref{eqn:SFC1_1}, the flexibility bid $\gamma$ was maximized, which is the primary remuneration in the Swiss AS market~\cite{qureshi2016economic}. As formulated in~\eqref{eqn:SFC_building}, \eqref{eqn:SFC1_3} used the DDP to predict the building temperature, incorporating constraints on comfort and the HP input. Similarly, by~\eqref{eqn:SFC_ESS}, \eqref{eqn:SFC1_3} used battery dynamics to prevent overcharging and overdischarging while satisfying the battery power bound. The total electrical power was required to track the baseline with a level of flexibility based on the AGC signal, as indicated in~\eqref{eqn:SFC1_4} and~\eqref{eqn:SFC1_5}. Finally, \eqref{eqn:SFC1_6} enforced a positive power flexibility $\gamma$.

In~\eqref{eqn:SFC1_5}, $\mathbf{\hat{P}}_{int}^{(j)}$ denotes the predicted intraday power transaction policy, which helps reduce conservatism. 
It was computed before solving~\eqref{eqn:SFC1}. For the steps $i=1,2,3$, $\mathbf{\hat{P}}_{int}^{(j)}(i)=0$. For steps $i=4,5,\dots, 96$ in the next day, $\mathbf{\hat{P}}_{int}^{(j)}(i)$ was computed by: 
\begin{equation}\label{eqn:SFC1_agc}
    \begin{aligned}
        \mathbf{\hat{P}}_{int}^{(j)}(i) &
        = \left| 
        \underset{\mathbf{\hat{P}}_{int}^{(j)}(i)}{\text{argmin}}
        \;\; 
        \overbrace{
            \sum_{k=1}^{i-4} \left[ \mathbf{\hat{P}}_{int}^{(j)}(k) +\bm{\alpha}_{scen}^{(j)}(k) \right]  }^{(a)}
            \right.\\
        & \left.
        + \overbrace{
            \sum_{k=i-3}^{i} \left[ \mathbf{\hat{P}}_{int}^{(j)}(k) + \frac{\sum_{l=1}^{N_{scen}}\bm{\alpha}_{scen}^{(l)}(k)}{N_{scen}} \right] }^{(b)}
            \right|
    \end{aligned}
\end{equation}
This equation utilized a causal policy to minimize the absolute accumulated sum of power transactions and AGC signals prior to the step $i$~\cite{fabietti2018multi}. Because transactions must be submitted at least 45 minutes in advance in the Swiss market, only the AGC signals before $i-3$ were known as shown in part (a). Part (b) used the average AGC scenarios to estimate the unknown signals.
For each step $i$, the previous $\mathbf{\hat{P}}_{int}^{(j)}(1),\dots,\mathbf{\hat{P}}_{int}^{(j)}(i-1)$ were computed via dynamic programming. Hence, the above problem was solved through straightforward arithmetic calculations. In summary, $\mathbf{\hat{P}}_{int}^{(j)}$ effectively eliminated large accumulated AGC biases, thus increasing the flexibility bid $\gamma$ for greater economic benefits.

%


    
    
    
    


\vspace{1.0em}

\subsubsection{Predictive building controller}
 

Starting each day at 00:00, the predictive building controller began its operations. It received the outputs ($\bar{P}$ and $\gamma$) from the day-ahead planner, optimized the HP power setpoint $u$ and the intraday power transaction $\bar{P}_{int}$ every 15 minutes. According to Swissgrid regulations, $\bar{P}_{int}$ can be adjusted 45 minutes before the actual AGC signal is assigned, improving the flexibility of AGC tracking. This predictive building controller was also developed from using the DDP and bi-level DeePC. Every 15 minutes, the controller executed the following steps repeatedly:



\vspace{1.0em}

\begin{itemize}[topsep=0em,itemsep=0.1em] 
    \item[1.] (This step is only conducted at 00:00 each day). Retrieve the recent operational data. If the two prerequisites~\eqref{eqn:PE_check} and~\eqref{eqn:PC_check} are met, update the Hankel matrix.
    \item[2.] Construct the $t_{init}$-step vectors $\mathbf{y}_{init}, \mathbf{u}_{init}, \mathbf{w}_{init}$ from the most recent operational data.
    \item[3.] Obtain the weather forecast from the weather API and create the $N_{}$-step vector $\bar{\mathbf{w}}_{pred}$.
    \item[4.] Predict the AGC signal, which will be described in Section~\ref{sect:V_para}, and construct the $N_{}$-step vector $\bar{\bm{\alpha}}_{pred}$.
    \item[5.] Retrieve the current \ac{soc} $SoC_{t}$ from the ESS.
    \item[6.] Solve the following robust optimization problem with a 15-minute resolution. The details of the control policy~\eqref{eqn:SFC2_6} will be described later.
\end{itemize}
\begin{subequations}\label{eqn:SFC2}
\begingroup
\allowdisplaybreaks
\begin{align}
    &\underset{\substack{  M_{w,i},M_{a,i},\mathbf{v}_i}}{\text{minimize}} \; && \; \sum_{i=1}^{N_{}} 
        \begin{array}{@{}c@{}} W_{u}\mathbf{v}_1(i)^2 + W_{P}\mathbf{v}_3(i)^2 + \\ W_{SoC}\left(\overline{\mathbf{SoC}}(i)-SoC_{ref}\right)^2 \end{array} \label{eqn:SFC2_1} \\
    &\text{s.t.} && \nonumber \\
    & \text{(Building system)} &&  \mathbf{u}_{pred} \in \mathbb{P}^H\left(\begin{array}{@{}c@{}} \mathbf{u}_{init},\mathbf{w}_{init},\\ \mathbf{y}_{init},\mathbf{w}_{pred}\end{array} \right) \label{eqn:SFC2_2} \\
    & \text{(ESS system)} &&  \mathbf{P}_{pred}^{E}  \in \mathbb{P}^E(SoC_{t}) \label{eqn:SFC2_3}\\
    & \text{(Total power)} && \mathbf{P} = \mathbf{P}_{pred}^{E} + \mathbf{u}_{pred} \label{eqn:SFC2_4} \\
    & \text{(Power tracking)} && \mathbf{P} = \mathbf{\bar{P}}+ \mathbf{P}_{int} + \gamma \bm{\alpha}_{pred} \label{eqn:SFC2_5} \\
    & \text{(Control policies)} && 
 \begin{bmatrix} \mathbf{u}_{pred} \\ \mathbf{P}^E_{pred} \\ \mathbf{P}_{int} \end{bmatrix}  =  \begin{bmatrix} M_{w,1} & M_{\alpha,1} \\ M_{w,2} & M_{\alpha,2} \\ M_{w,3} & M_{\alpha,3} \end{bmatrix} \begin{bmatrix} \mathbf{w}_{pred} \\ \bm{\alpha}_{pred} \end{bmatrix} \nonumber \\
 & && \quad\quad + \begin{bmatrix} \mathbf{v}_1^{\top} & \mathbf{v}_2^{\top} & \mathbf{v}_3^{\top} \end{bmatrix}^{\top} \label{eqn:SFC2_6} \\
    & \text{(Disturbances)}&& \begin{array}{@{}l@{}l@{}}    & \forall \mathbf{w}_{pred}  \in \bar{\mathbf{w}}_{pred} \oplus \mathbb{W}, \\ & \forall \bm{\alpha}_{pred} \in \bar{\bm{\alpha}}_{pred} \oplus \mathbb{A}   \end{array} \label{eqn:SFC2_7}   
\end{align}
\endgroup
\end{subequations}  
   
\begin{itemize}
    \item[7.] Retrieve the optimal solution $\mathbf{u}_{pred}$ and $\mathbf{P}_{int}$ of ~\eqref{eqn:SFC2}. Send the modulated signal based on the power setpoint with $u:=\mathbf{u}_{pred}(1)$ to the building BMS and submit the 45-minute ahead transaction with $P_{int,t+3} $ $= \mathbf{P}_{int}(4)$ to the ESS controller.
    \item[8.] Stay idle until the next iteration.
\end{itemize}

The details of the predictive building controller are discussed below, including the objective, constraints and the worst-case robust programming.
The objective~\eqref{eqn:SFC2_1} aimed to minimize several nominal costs based on the nominal control variables $\mathbf{v}_i$. The term $\mathbf{v}_1(i)^2$ was minimized to reduce the building's energy consumption. The cost $\mathbf{v}_3(i)^2$ sought to limit power transactions to necessary levels, as extreme power transactions could disrupt power tracking due to the imperfect dynamic predictions and AGC signal forecasts. Furthermore, $\overline{\mathbf{SoC}}$ represents the nominal \ac{soc} predicted by the ESS dynamics and $\mathbf{v}_2$. By minimizing the difference  between this term and $SoC_{ref}:= \frac{SoC{max}+SoC_{min}}{2}$, the battery state was kept around the median value, enhancing the reliability of power tracking and reducing penalties. The cost weights $W_{u}, W_{P}, W_{SoC}$ were adjusted based on user priorities and objectives.

Regarding the constraints, \eqref{eqn:SFC2_2} leveraged the DDP to predict the future $N$ steps for the building, including conditions on indoor temperature comfort and the HP input limitations. The constraint~\eqref{eqn:SFC2_3} predicted ESS states, ensuring that the operational constraints were met.
The total electrical power, according to \eqref{eqn:SFC2_4} and~\eqref{eqn:SFC2_5}, tracked the sum of the power baseline, intraday power transaction, and some flexibility based on the predicted AGC signal. Unlike~\eqref{eqn:SFC1_5}, $\mathbf{\bar{P}}$ and $\gamma$ were fixed by the day-ahead planner.

The constraint~\eqref{eqn:SFC2_7} integrated worst-case robust programming to account for the prediction errors related to weather conditions and AGC signals. This approach was designed to improve power tracking robustness and reduce potential penalties in SFC tasks. 
\changeD{
In addition, the control policies in~\eqref{eqn:SFC2_6} used a casual structure to reduce conservatism in robust programming. In detail, the control policies for $\mathbf{P}_{pred}^{E}, \mathbf{u}_{pred}$ were defined as:
}
\begin{align*}
             \mathbf{v}_i^\top :&= \begin{bmatrix} (v_i^{1}) & \cdots & \cdots & (v_i^{N_{})^\top}\end{bmatrix}^\top, \\
             M_{\beta,i} :&= \begin{bmatrix}
                0_{n_i\times n_{\beta}} & \cdots & \cdots & 0_{n_i\times n_{\beta}} \\
                M_{\beta,i}^{2,1} & 0_{n_i\times n_{\beta}} & \ddots & 0_{n_i\times n_{\beta}} \\
                \vdots & \ddots & \ddots & \cdots \\
                M_{\beta,i}^{N_{},1} & \cdots & M_{\beta,i}^{N_{},N_{}-1} & 0_{n_i\times n_{\beta}} \\
            \end{bmatrix}, \\
             \text{for:}\; & \beta \in \{w, \alpha\},  i \in \{1,2\}; n_{1} = n_u, \; n_{2} = 1, 
 \end{align*}
 Here, some values were forced to be zero, while $M_{\beta,i}^{\cdot,\cdot}$ and $v_i^{\cdot}$ were decision variables. This structure made the input at time $t$ an affine function of the previous disturbance up to time $t-1$. 
 This causal structure leverages the information of the closed-loop disturbances, leading to better performance than open-loop robust MPC~\cite{goulart2006optimization, oldewurtel2013stochastic}. Moreover, the affine structure maintains convexity, ensuring the problem remains tractable. Similarly, the control policy for $\mathbf{P}_{int}$ was structured as,
\begingroup
\allowdisplaybreaks
\begin{align*}
\mathbf{v}_3 :&= \begin{bmatrix} P_{int,t} &P_{int,t+1} & P_{int,t+2} & v_3^{4} & \cdots & \cdots & v_3^N \end{bmatrix}^\top, \\
             M_{\beta,3} :&= \begin{bmatrix}
                0_{4\times n_{\beta}} & \cdots & \cdots & 0_{4\times n_{\beta}}\\
                M_{\beta,3}^{5,1} & 0_{1\times n_{\beta}} & \ddots & 0_{1\times n_{\beta}} \\
                \vdots & \ddots & \ddots & \cdots \\
                M_{\beta,3}^{N_{},1} & \cdots & M_{\beta,3}^{N_{},N_{}-1} & 0_{1\times n_{\beta}} \\
            \end{bmatrix}, \\
            & \text{for:}\; \beta \in \{w, a\}
 \end{align*}
 \endgroup
where $P_{int,t},P_{int,t+1}$, and $P_{int,t+2}$ were the power transactions determined before time $t$. 
\changeD{
At time $t$, $P_{int,t+3}$ was determined by solving for $v_3^{4}$ due to the 45-minute ahead transaction requirement.
In summary, this controller enabled the flexible operation of the building system while keeping robustness against various disturbances and prediction errors.
}


\vspace{-0.5em}
\begin{remark}
The DDP and bi-level DeePC bypass explicit modeling and state estimation steps, which are commonly required for MPC with a parametric model.
Due to the linearity of the DDP, both optimization problems~\ref{eqn:SFC1} and~\ref{eqn:SFC2} are convex and can be solved efficiently. 
In detail, \eqref{eqn:SFC1} is a \ac{lp}, which can be quickly solved by LP solvers such as~\textit{CLP} and~\textit{MATLAB Linprog}.
In~\eqref{eqn:SFC2}, the objective function~\eqref{eqn:SFC2_1} is quadratic, while the constraints~\eqref{eqn:SFC2_2} to \eqref{eqn:SFC2_6} are linear with respect to the disturbances $\mathbf{w}_{pred}$ and $\bm{\alpha}_{pred}$. If polytopic disturbance constraints are used in~\eqref{eqn:SFC2_7}, the entire problem can be reformulated into a QP using dualization techniques. There are many efficient QP solvers and robust optimization toolboxes available, such as \textit{MOSEK} and \textit{YALMIP}~\cite{lofberg2004yalmip}. 
\end{remark}

\vspace{-1.0em}
\begin{remark}
    Occasionally, both optimization problems~\ref{eqn:SFC1} and~\ref{eqn:SFC2} became infeasible. In climate control, for instance, this usually happens at noon when the indoor temperature reaches its upper bound and predicted outdoor temperatures exceed the cooling capacity. In the case of infeasibility, the box constraints on $y$ and $SoC$ were softened by introducing slack variables. 
    These variables were optimized to be small by adding a significant penalty cost to the objective function. 
    This approach ensures feasibility while preserving satisfactory performance in practice.
\end{remark}

\begin{remark}
\changeC{
To deliver SFC services, similar hierarchical controllers have been developed and refined in previous work~\cite{qureshi2016economic,gorecki2017experimental,fabietti2018multi}. However, there are two key differences between the setups in this work and the prior research. First, previous studies relied on fixed state-space models derived from system identification for short-term experiments, whereas this work used an adaptive DDP for longer-term tests. Second, the building controller in this study was designed to be robust to predictive uncertainty, in contrast to the certainty-equivalent controllers used in earlier work. 
}
\end{remark}



 
    


\subsubsection{ESS controller}
The ESS controller adjusted the input $P^E$ of the ESS to track the real-time baseline, based on the power baseline, the AGC signals, and HP's power. The controller executed the following sequence of steps with a 4-second sampling period:


\begin{itemize} 
    \item[1.] (Only executed at 00:00 daily) The power flexibility parameter $\gamma$ and the day-ahead power baseline $\mathbf{\bar{P}}$ are updated.
    \item[2.] If the predictive building controller is running, it receives the new $P_{int,t}$. If not, $P_{int,t} = P_{int,t-1}$ is used.
    \item[3.] The actual power measurements $P_{t}^H$ of the HP are retrieved. (Note that there might be slight deviations from the setpoint $u$ calculated by the predictive building controller.)
    \item[4.] The setpoint $P_t^E$ is calculated using the equation below, and the ESS is accordingly controlled:
        \begin{equation}
        \begin{aligned}
        P_t^{E} = \bar{P}_{t} + P_{int,t} + \gamma \alpha_t - P_t^{H}
        \end{aligned}
        \end{equation} 
    \item[5.] Stay idle until the next iteration
    \vspace{-0.5em}
\end{itemize}

\section{Results: prediction tests and DR experiments } \label{sect:V}
\changeC{
This section presents the results of the prediction tests for DDP and the DR experiments conducted at EPFL. 
Section~\ref{sect:III} shows the effectiveness of the adaptive DDP prediction and its tuning process using the operational data from the Polydome.
Based on these tuning results, we outline the parameter selection for the SFC controller in Section~\ref{sect:III_tune}. In Section~\ref{sect:V_result}, we analyze the experimental results, including daily operations, indoor temperature statistics and daily costs, followed by a comparative analysis. The collected evidence strongly supports the efficiency and efficacy of our proposed DDP and bi-level DeePC method.
}

\subsection{Practical insights for prediction with DDP } \label{sect:III}
\changeC{
In this subsection, we conducted the prediction tests for the DDP using the historical operational data from the Polydome. Section~\ref{sect:III_ada} illustrates the improved prediction performance of the adaptive DDP compared to DDP with a fixed Hankel matrix. Additionally, we performed a sensitivity analysis of DDP's hyperparameters, demonstrating that it required less tuning effort than the standard DeePC.
}


\subsubsection{Efficacy of the Hankel matrices update} \label{sect:III_ada}


We analyzed the predictive performance of DDP using both adaptive and fixed Hankel matrices, with data collected from the Polydome between June and September 2021. During this period, indoor temperature control was managed by a default controller, randomized control inputs, or a robust bi-level DeePC controller. The dataset can be found can be found on GitHub\footnote{\href{https://github.com/PREDICT-EPFL/polydome}{https://github.com/PREDICT-EPFL/polydome}}. Table~\ref{tab:deepc_prediction} summarizes the prediction performance statistics for various prediction horizons
The data were partitioned into three segments: an initial Hankel matrix construction phase (using the first 5 days for $N=12$ or $N=24$, and 10 days for $N=48$ or $N=96$), a 10-day validation set for parameter tuning, and the remaining data for prediction testing. From Table~\ref{tab:deepc_prediction}, it is evident that the adaptive updating method consistently outperformed the fixed approach in all the cases.

\begin{figure}[!ht]
    \centering \includegraphics[width=1.0\linewidth]{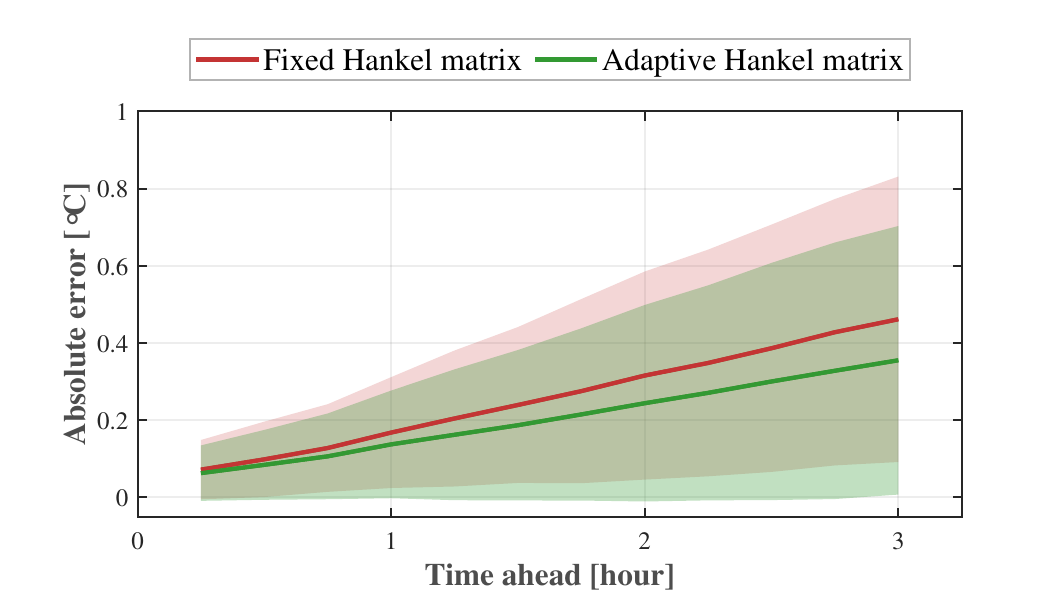}
    \caption{Absolute error distribution in the prediction horizon by mean and standard deviation: 12-step prediction (3 hours) }
    \label{fig:pdf_deepc_pred_N12}
\end{figure}

\begin{figure}[!ht]
    \centering \includegraphics[width=1.0\linewidth]{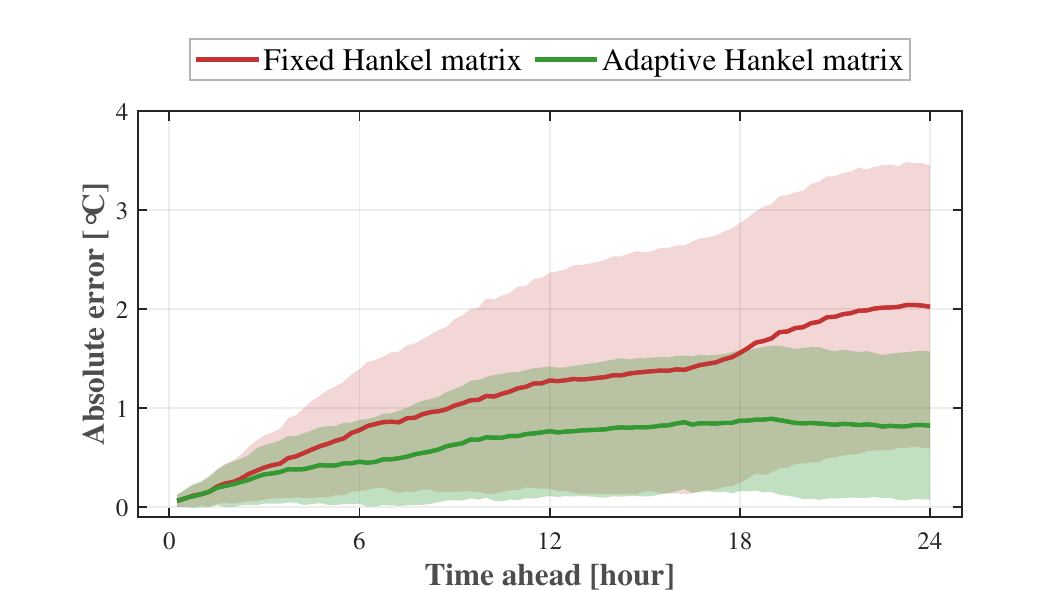}
    \caption{Absolute error distribution in the prediction horizon by mean and standard deviation: 96-step prediction (24 hours) }
    \label{fig:pdf_deepc_pred_N96}
\end{figure}

Since the parameters $N=12$, $N=96$ were used in our SFC case study (explained in Section~\ref{sect:V_result}), we present a more detailed comparison in Figures~\ref{fig:pdf_deepc_pred_N12} and~\ref{fig:pdf_deepc_pred_N96}. These figures provide display the \ac{mae} and standard deviation relative to the prediction horizon. The adaptive method consistently showed a smaller absolute error compared to the fixed approach. The performance gain from the adaptive matrix update grew as the prediction horizon increased, particularly in the $N=96$ scenario. Beyond a 12-hour prediction window, the fixed approach's MAE surpassed 1$\degree C$ and gradually rose to 2$\degree C$. In contrast, the adaptive approach kept the MAE below 1$\degree C$ throughout the entire prediction horizon.

\begingroup
\renewcommand{\arraystretch}{1.5} 
\begin{table}[!ht]
\centering
\footnotesize
\begin{tabular}{  b{1cm}  b{1cm} b{1cm} b{1cm} b{1cm} b{1cm}} 
  \hline
  Prediction  & Hours & \multicolumn{2}{c}{Validation set} & \multicolumn{2}{c}{Test set} \\ 
  steps & ahead & Fixed & Adaptive & Fixed & Adaptive \\
  \hline
  12  & 3 & 0.279 & \textbf{0.207} & 0.258 & \textbf{0.203}\\ 
  24  & 6 & 0.572 & \textbf{0.285} & 0.522 & \textbf{0.357}\\ 
  48  & 12 & 0.608 & \textbf{0.434} & 0.763& \textbf{0.419}\\ 
  96  & 24 & 1.050 & \textbf{0.574} & 		1.194 & \textbf{0.652}\\   
  \hline  
\end{tabular}
\caption{Comparison of the \ac{mae} over different prediction horizons} 
\label{tab:deepc_prediction}
\end{table}
\endgroup

\subsubsection{Parameter sensitivity analysis} \label{sect:III_tune}
\changeD{
The DDP's tuning parameters include the regularization cost weight $\mathcal{E}g$, the data length for the Hankel matrix $T$, and the initial steps count $t_{init}$. Unlike the standard DeePC~\cite{coulson2019data,chinde2022data}, which requires closed-loop tests, the DDP can be tuned using historical data. We conducted a sensitivity analysis based on the two setups used in the SFC case study (i.e. N=12 and N=96) using the same historical operational data in Section~\ref{sect:III_ada} for adaptive DDP prediction.
}



\begin{figure}[!ht]
    \centering \includegraphics[width=1.0\linewidth]{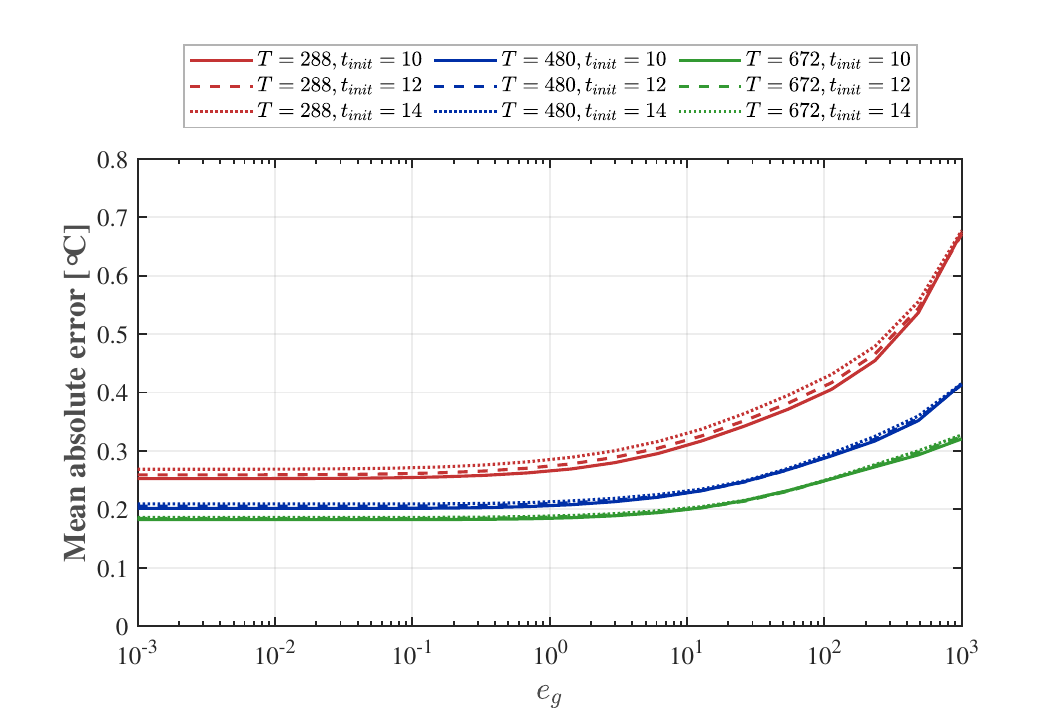}
    \caption{MAE sensitivity of $\mathcal{E}_g = e_g I$ in DeePC prediction: 12-step prediction (3 hours) }
    \label{fig:pdf_deepc_sen_N12}
    \vspace{-1.5em}
\end{figure}

\begin{figure}[!ht]
    \centering \includegraphics[width=1.0\linewidth]{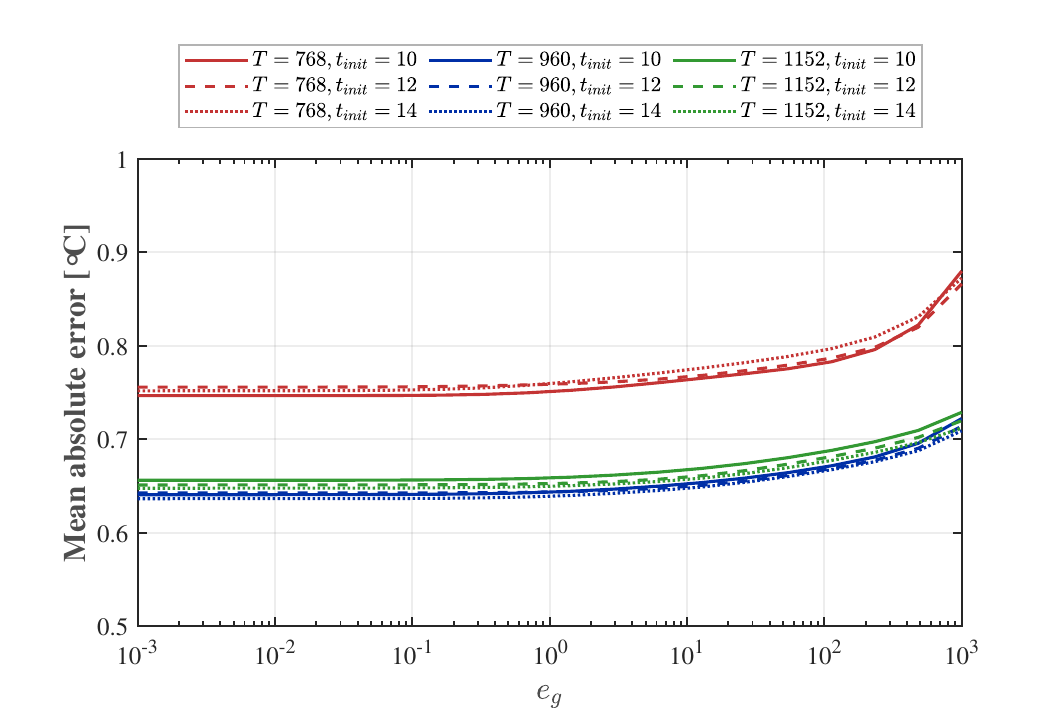}
    \caption{MAE sensitivity of $\mathcal{E}_g = e_g I$ in DeePC prediction: 96-step prediction (24 hours) }
    \label{fig:pdf_deepc_sen_N96}
    \vspace{0em}
\end{figure}

We started with a sensitivity analysis on the regularization weight $\mathcal{E}_g$, by selecting $\mathcal{E}_g = e_g I$ and performing a grid search on $e_g$. Figure~\ref{fig:pdf_deepc_sen_N12} presents the results for the N=12 instance, showing the MAE across different $e_g$ values. Each curve corresponds to a specific set of $\{T, t_{init}\}$ options. Notably, a wide range $e_g\in [10^{-3},10]$ produced consistent prediction performance. When $\mathcal{E}_g$ was too large, the prediction accuracy decreased. Similar results were observed for $N=96$ in Figure~\ref{fig:pdf_deepc_sen_N96}. Importantly, $\mathcal{E}_g$ exhibited low sensitivity to variations in $T$ and $t_{init}$, and the adaptive data updates, which is critical for long-term reliable operation. If instead $\mathcal{E}_g$ were sensitive to these adaptive DDP updates, it would complicate the use of DDP for different purposes. 

\begin{figure}[!ht]
    \centering \includegraphics[width=1.0\linewidth]{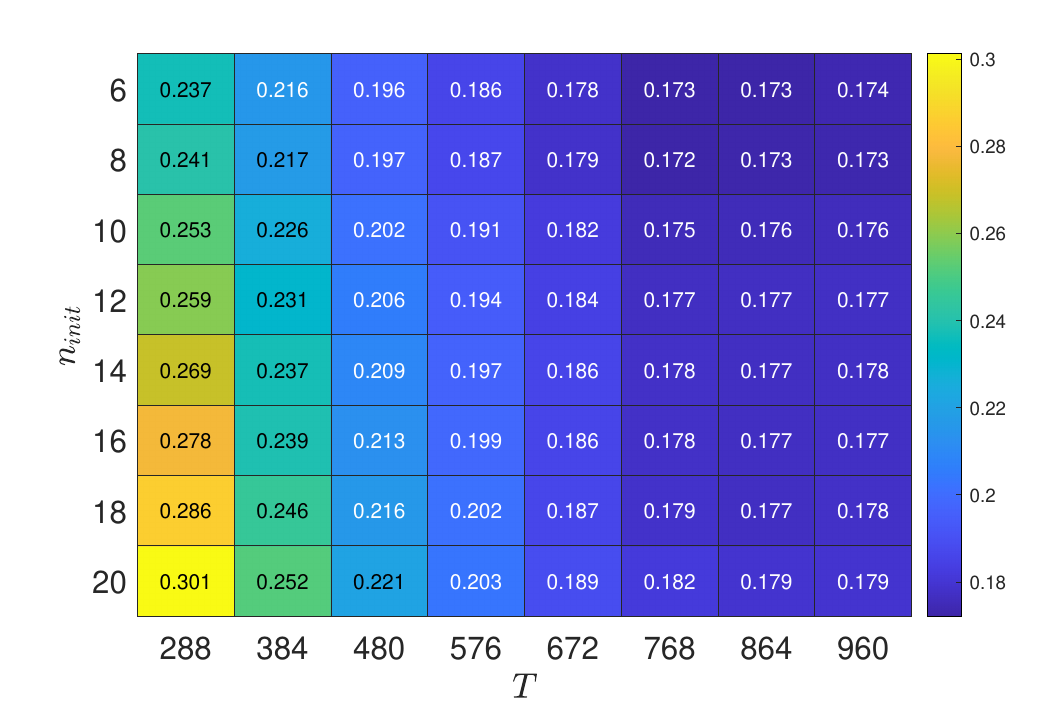}
    \caption{MAE sensitivity of $T$ and $t_{init}$ in DeePC prediction: 12-step prediction (3 hours), $\mathcal{E}_g=0.01I$ }
    \label{fig:pdf_deepc_sen_N12_2}
    \vspace{-1.5em}
\end{figure}

\begin{figure}[!ht]
    \centering \includegraphics[width=1.0\linewidth]{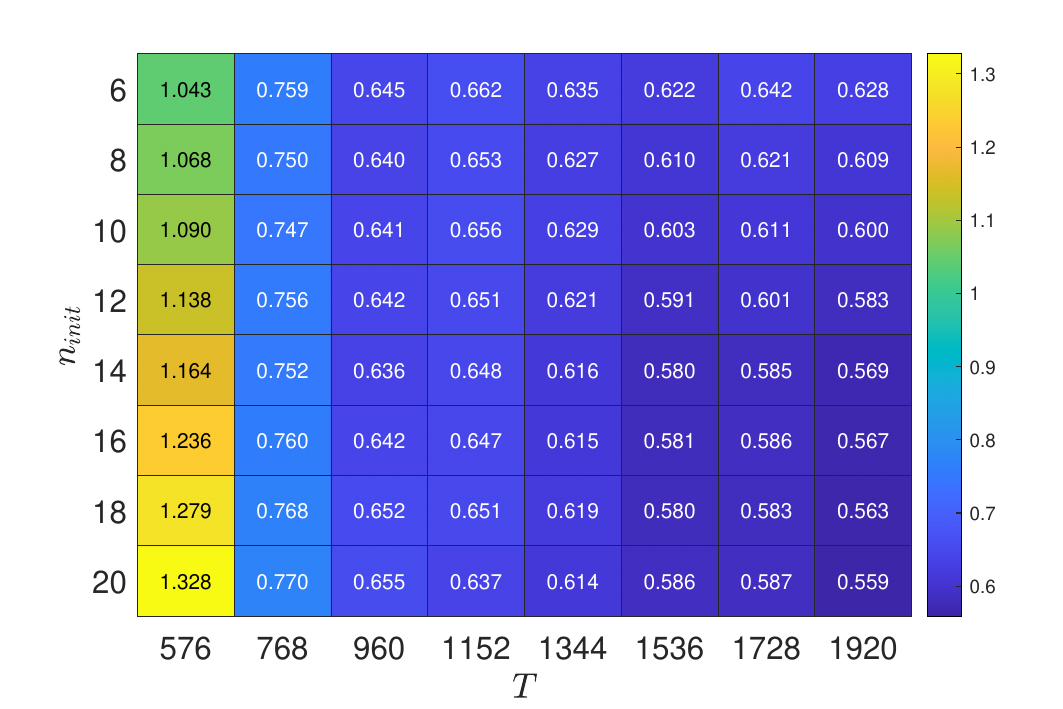}
    \caption{MAE sensitivity of $T$ and $t_{init}$ in DeePC prediction: 96-step prediction (24 hours), $\mathcal{E}_g=0.01I$ }
    \label{fig:pdf_deepc_sen_N96_2}
    \vspace{0em}
\end{figure}

Next, we conducted a sensitivity analysis on $T$ and $t_{init}$, keeping $\mathcal{E}g$ constant at $0.01I$ based on the findings above. Figure~\ref{fig:pdf_deepc_sen_N12_2} displays a heatmap of the MAE for various $\{T, t_{init}\}$  with $N=12$. As darker shades of blue indicate lower MAE values, it indicates that prediction performance exhibited relatively insensitive to the tuning parameters. Specifically, increasing $T$ while keeping $t_{init}$ constant led to small but consistent improvements in MAE. For data sampled every 15 minutes, a ten-day dataset was generally sufficient for both short-term and long-term predictions. Furthermore, Figure~\ref{fig:pdf_deepc_sen_N12_2} reveals that raising $t_{init}$ did not necessarily enhance prediction outcomes; in fact, it could even reduce the performance when data was limited. However, for larger $T$ values, the impact of $t_{init}$ on the MAE diminished. Figure~\ref{fig:pdf_deepc_sen_N96_2}, depicting results for $N=96$, reflects similar trends to those observed for $N=12$, with a slight variation: larger $t_{init}$ values yielded better prediction outcomes when $T$ was sufficiently large.

In summary, the results suggest choosing $\mathcal{E}_g = e_g I$  with a small $e_g$ value (i.e. $e_g < 1 $) for the case study. While a larger data size $T$ can enhance prediction accuracy, a ten-day dataset is generally sufficient. As for $t_{init}$, when $T$ is sufficiently large, it is prudent to choose an overestimated state order for $t_{init}$.


\subsection{Choice of parameter for experiments} \label{sect:V_para}
Table~\ref{tab:para} summarizes the choices for the parameters. For the number of prediction steps $N$, $N=96$ was used in the day-ahead planner to match the 15-minute sampling time. In the predictive building controller, we selected $N=12$ due to DDP's good prediction performance over this horizon, which helped avoid excessive conservatism in the robust optimization.
The DDP parameters were chosen based on the tuning result from Section~\ref{sect:III_tune}.

\begin{table}[!ht]
\footnotesize
\begin{center}
\begin{tabular}{  l  l}
  \hline
  Day-ahead planner & \\ 
  \hline
  No. of Prediction Steps $N_{}$ & 96 \\ 
  No. of Initial Steps in DeePC $t_{init}$ & 12\\ 
  Length of Operational data in DeePC $T$ & 960\\ 
  Regularization weight $\mathcal{E}_g$ & 0.01$I$ \\  
  Number of AGC scenarios $N_{scen}$ & 300\\ 
  \hline  
  Predictive building controller & \\ 
  \hline  
  No. of Prediction Steps $N_{}$ & 12\\ 
  No. of Initial Steps in DeePC $t_{init}$ & 12\\ 
  Length of Operational data  in DeePC $T$ & 480\\ 
  Regularization weight $\mathcal{E}_g$ & 0.01$I$ \\  
  Error Bound of the Weather Forecast  $\mathbb{W}$ & 
  $ |e_w|  \leq  \begin{bmatrix} 0.2\celsius \\ 0.05kW/m^2  \end{bmatrix} $ \\
  Error Bound of the AGC Signals  $\mathbb{A}$ & 
  $ |e_a|  \leq 0.2 $ \\
  \hline  
  Building and ESS constraints & \\
  \hline  
  Indoor temperature bounds $\left[y_{min}^H, y_{max}^H\right]$ & $\left[22\celsius, 26\celsius\right]$\\
  HP input bounds $\left[u_{min}^H, u_{max}^H\right]$ & $\left[2.4kW, 8.4kW\right]$ \\
  ESS SoC bounds $\left[SOC_{min}^S, SOC_{max}^S\right]$ & $\left[0.25kWh, 5kWh\right]$\\
  ESS input bounds $\left[P_{min}^S, P_{max}^S\right]$ & $\left[-5kW, 5kW\right]$\\
  \hline 
\end{tabular}
\caption{Choice of parameters for the experiment} \label{tab:para}
 \end{center}
\end{table}

There were additional parameters associated with the SFC controllers~\eqref{eqn:SFC1} and~\eqref{eqn:SFC2}. We incorporated 300 historical AGC scenarios from the year 2019, sourced from Swissgrid. The error bound for the AGC signals, denoted as $\mathbb{A}$, was derived from an AGC forecasting process using a non-parametric probabilistic forecasting method~\cite{pinson2009probabilistic}. We implemented a forecast test using the 2019 AGC signal to establish $\mathbb{A}$. Moreover, we retrieved weather forecasts and actual weather data from the weather API, \textsc{Tomorrow.io}~\cite{tomorrow}, and established the error bound $\mathbb{W}$.

The indoor comfort range was determined following the European Standard~\cite{cen200715251} for classrooms. The constraints for the \ac{hp} and \acp{ess} systems were established based on their physical limitations, as outlined in Section~\ref{sect:IV_testbed}. The lower bound of the ESS's \ac{soc} was set to be $0.25kWh$ to prevent over-discharging.

\subsection{Experiment results} \label{sect:V_result}


This section reports the experiment conducted at the Polydome test bed from July 14th, 2022 to September 3rd, 2022. Over this 52-day period, the Polydome and the ESS were managed by the SFC controller to provide the SFC service within a simulated Swiss AS market.

The historical AGC signals from 2020, provided by Swissgrid, were used to emulate the AS market. Each day, the AS market emulator picked a random day from 2020 and sent the corresponding AGC signals to the SFC control module every 4 seconds. At 23:45 of the preceding day, the day-ahead planner calculated the bid for power flexibility and baseline.
In the day-ahead planner, we utilized only the cooling-mode I/O data in the DDP since mode switching was not controllable. To mitigate overestimation of flexibility, an additional cost term, $\sum_{j=1}^{N_{scen}}W_{base}\lVert\mathbf{P}^{(j)}-\mathbf{P}_{pre}\rVert_2^2$, was added to the objective function. Here, $\mathbf{P}_{pre}$ is the power baseline of the previous day, and $W_{base}$ is an appropriate weight.

Throughout the day, the predictive building controller set the HP's input and intraday power transactions every 15 minutes, while the ESS controller managed the battery at 4-second intervals
In our testbed, the \ac{hp} occasionally switched to heating mode at night even in summer. As the heating and cooling modes of the HVAC system exhibit different responses, we employed two distinct I/O datasets from each mode to build the DDP in the predictive building controller. Every day, the Hankel matrix for the cooling mode was updated using the latest collected data. Conversely, the Hankel matrix for the heating mode was built using data from the preceding winter and remained unchanged during summer, as the heating input of HP typically remained at 0.

The SFC controller, the simulated ESS system and the AS market were implemented in \textsc{MATLAB} on the central computer. The two optimization problems~\eqref{eqn:SFC1} and~\eqref{eqn:SFC2} were solved using \textsc{GUROBI}. The following sections detail the experimental results and compare the proposed method with the default building controller.

\subsubsection{Operation details}

\begin{figure*}[!ht]
    \centering
    \includegraphics[width=0.9\linewidth]{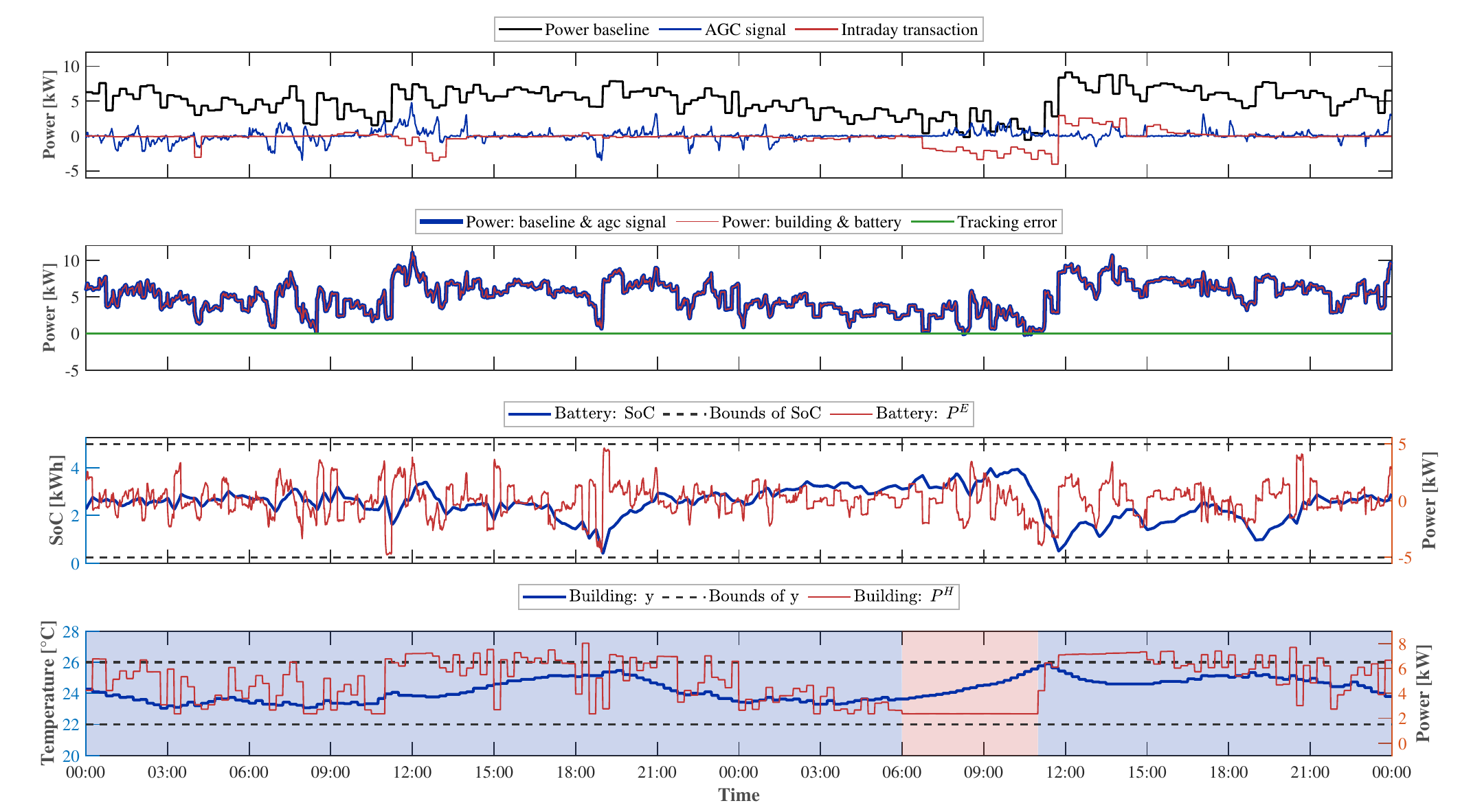}
    \caption{Operational details on 20$^{th}$ and 21$^{st}$ July, 2022.}
    \label{fig:pdf_day_details_1}
\end{figure*}

\begin{figure}[!ht]
    \centering
    \includegraphics[width=1.0\linewidth]{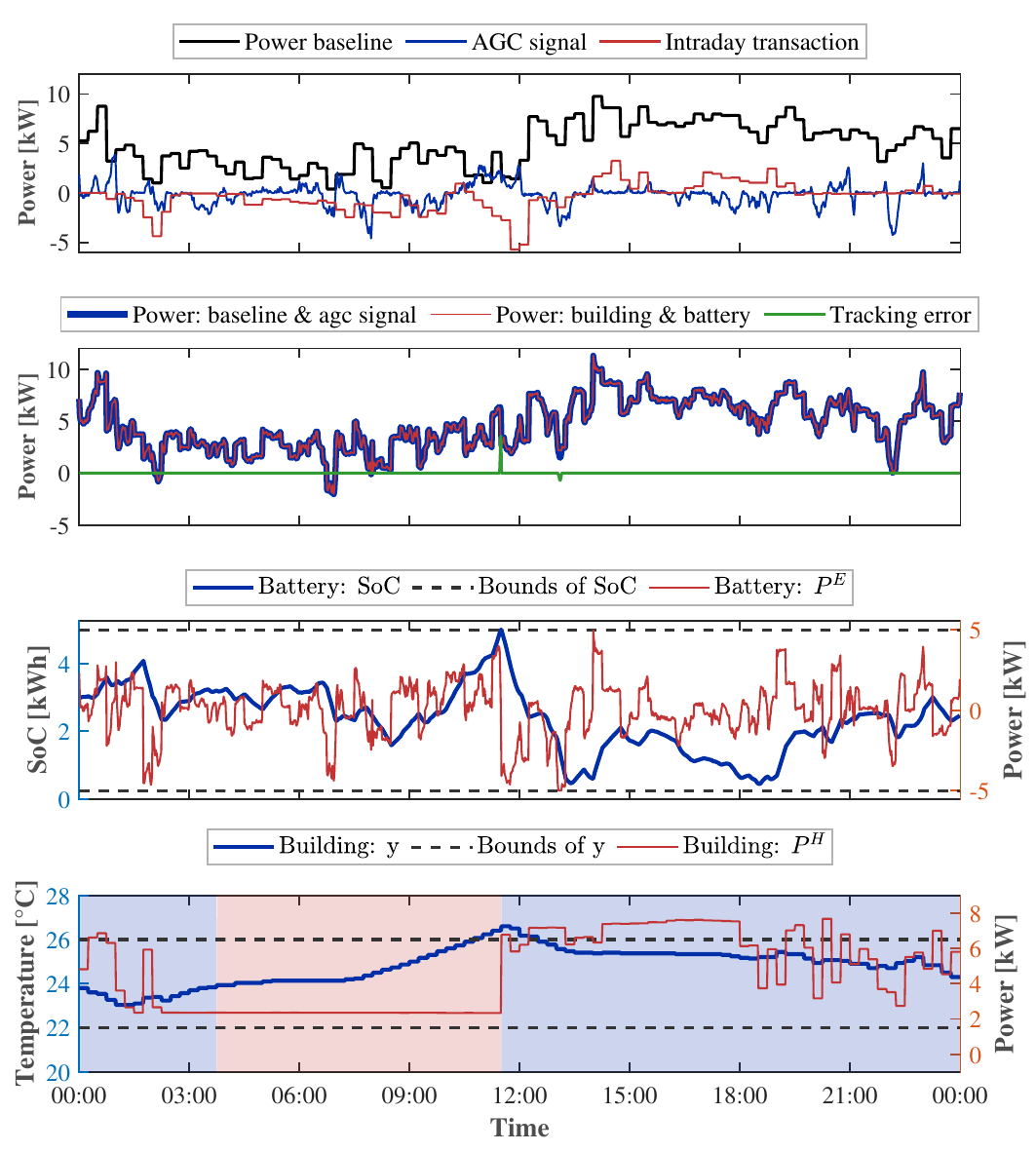}
    \caption{Operation details on 25$^{th}$ July, 2022.}
    \label{fig:pdf_day_details_2}
\end{figure}

\begin{figure}[!ht]
    \centering
    \includegraphics[width=1.0\linewidth]{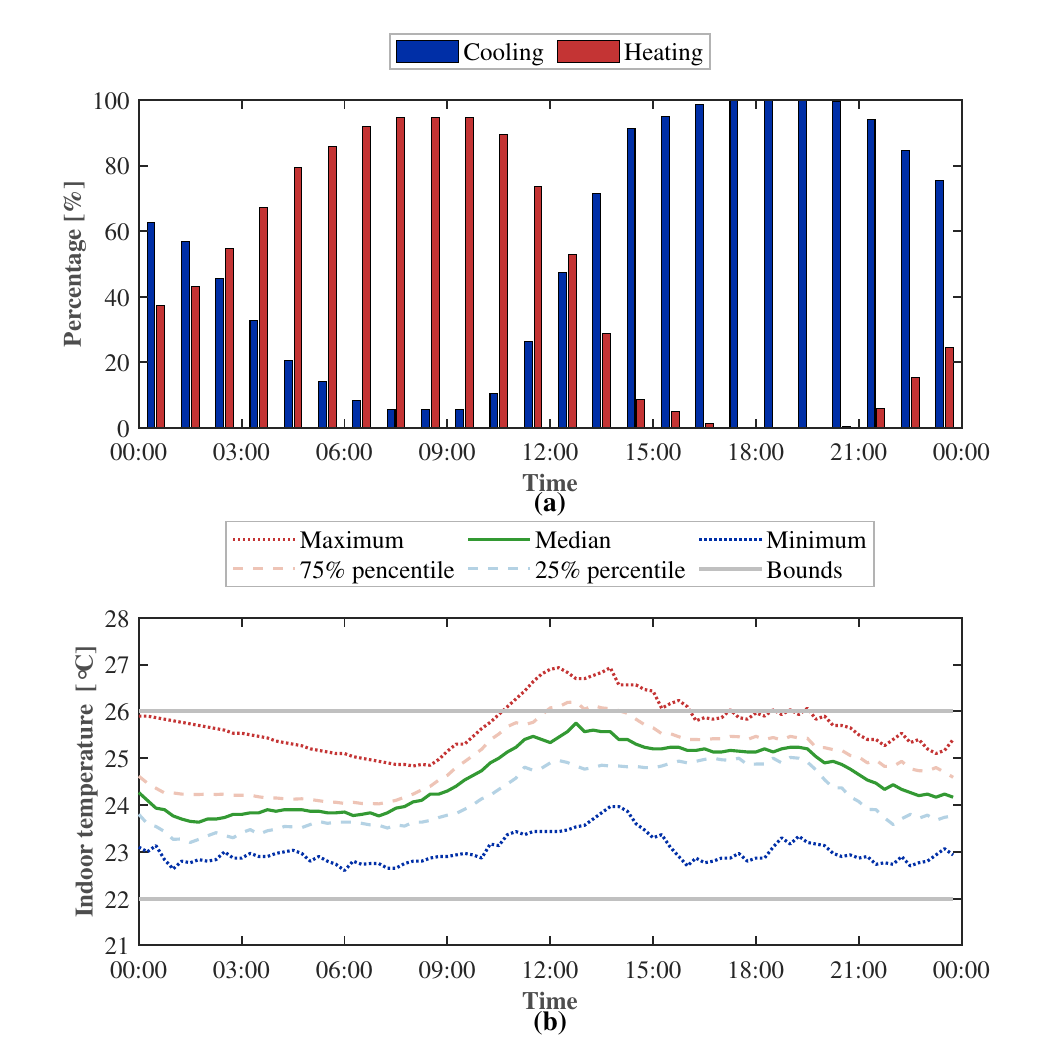}
    \caption{Statistics of the building operation: (a) the percentage of heating and cooling for each hour, (b) the distribution of indoor temperature.}
    \label{fig:pdf_build_distrbution}
\end{figure}

\begin{figure*}[!ht]
    \centering
    \includegraphics[width=1.0\linewidth]{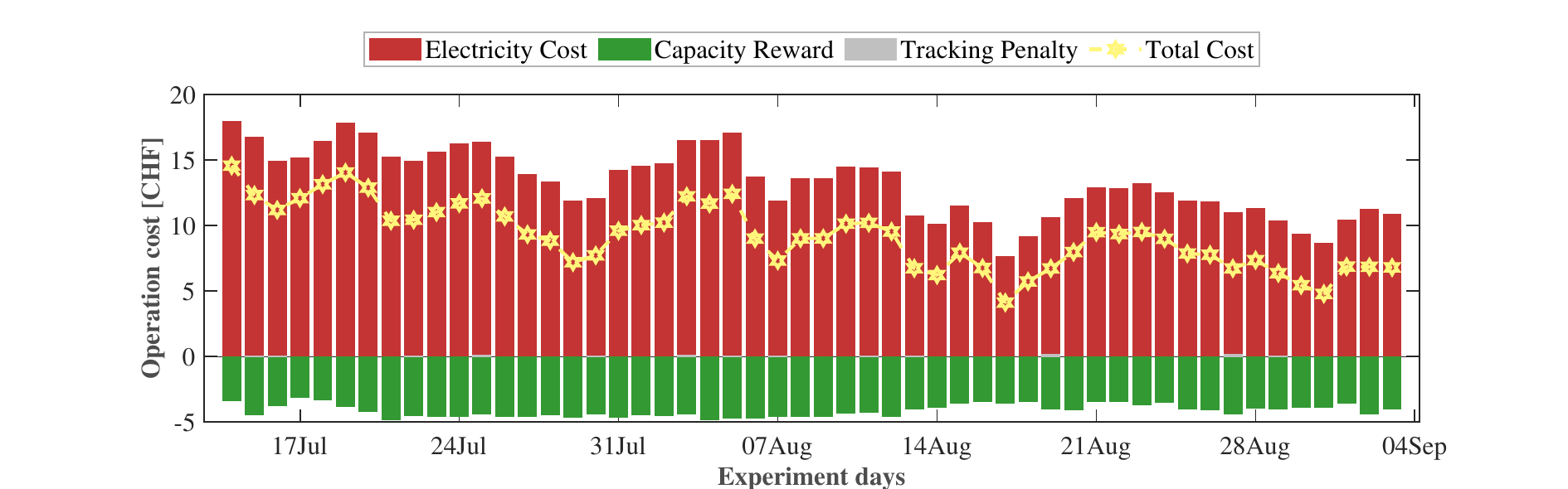}
    \caption{Different operating costs and the DR income of a continuous 57-day experiment: the battery and the building operate together using DeePC to support SFC.}
    \label{fig:pdf_cost_DR}
\end{figure*}

Figure~\ref{fig:pdf_day_details_1} depicts the operational specifics for two days - July 20th and 21st, 2022. The first subplot illustrates the original power baseline, intraday power transactions, and the AGC signals. The second subplot shows that the sum of building and ESS powers consistently aligned with the sum of the power baseline, intraday power transactions, and AGC signals, as evidenced by the tracking error (the green line) remaining at zero.
The lower two subplots provide detailed operational statuses of the Polydome and the ESS. The translucent blue and red colors indicate when the HP is in cooling and heating modes, respectively. Please note that the building's power consumption remained positive due to the continuous operation of the ventilation system, which consumed approximately 2.4 kW. The results show that the ESS responded quickly to fluctuations in the AGC signals, and the building's substantial energy capacity helped reset the battery when necessary. For instance, around 19:00 on the first day, the predictive building controller reduced power consumption when the ESS's SoC approached its lower limit.
Therefore, the SFC controller facilitated the effective operation of the Polydome and the ESS, ensuring adherence to both battery and HP's physical constraints while maintaining an indoor temperature within a comfortable range.



Mode switching was controlled by a scheduler within the BMS, as detailed in Section~\ref{sect:IV_testbed}. However, this scheduler occasionally produced mis-configured outcomes. Lausanne's significant temperature variations during the summer, where nighttime temperatures in August sometimes dropped below 15 $\celsius$, could trigger the HP's heating mode at night. As daytime temperatures rose, the scheduler's delay in switching modes could result in the HP continuing in heating mode even at midday. Since future mode switches were unknown, the SFC controller sometimes resulted in suboptimal indoor climate control. For instance, on July 25th, 2022, the indoor temperature exceeded the comfort level at 11:00 because the HP remained in the heating mode, as displayed in Figure \ref{fig:pdf_day_details_2}.


Despite these challenges due to mode switching, the predictive building controller mitigated the negative impact on system operation and intraday power tracking. As shown in Figure \ref{fig:pdf_day_details_2}, when in the heating mode and the future temperatures were predicted to be too high, the controller limited the HP  to ventilation only. Intraday power tracking was maintained as much as possible, with only a short-time violation caused by continuous positive AGC signals around 11:00.
Once the mode switched to cooling, the HP was instructed to start cooling, rapidly reducing the room temperature. 
Although the significant cooling input led to substantial ESS discharge and a tracking error due to a sudden AGC drop around 13:00, the robust controller quickly corrected the situation. It slightly reduced HP's power consumption, reset the ESS, and eliminated further tracking errors.

To provide a comprehensive view of the impact of mode switching throughout the experiment, we analyze the results from two perspectives: comfort and costs. First, Figure ~\ref{fig:pdf_build_distrbution} illustrates the 24-hour distribution of HP mode conditions and room temperature. Before 13:00, the heating mode was predominant, resulting in temperatures that exceeded the upper constraint. However, once the cooling mode was activated, the SFC ensured a rapid return to the comfortable temperature range.
Second, we present the daily cost distribution in Figure~\ref{fig:pdf_cost_DR}. Here, red represents the power used by the HP and ESS, green represents the reward obtained through flexibility bidding, gray represents penalties due to tracking errors, and yellow represents the total cost. All economic components were calculated based on the average cost data of August 2020 from Swissgrid. Although tracking errors occurred on some days, primarily due to mode changing and deviations in the AGC signal from predictions, these errors were minor compared to other factors. 

\changeC{
Overall, the DDP-based SFC controller demonstrated robust performance, effectively operating the entire system and successfully completing the SFC task, even under imperfect equipment settings.
To further assess the cost benefits of this SFC control, we compared it with the costs incurred by the default controller in the following section.
}

\subsubsection{Comparison}

\begin{figure}[!ht]
    \centering    \includegraphics[width=1.0\linewidth]{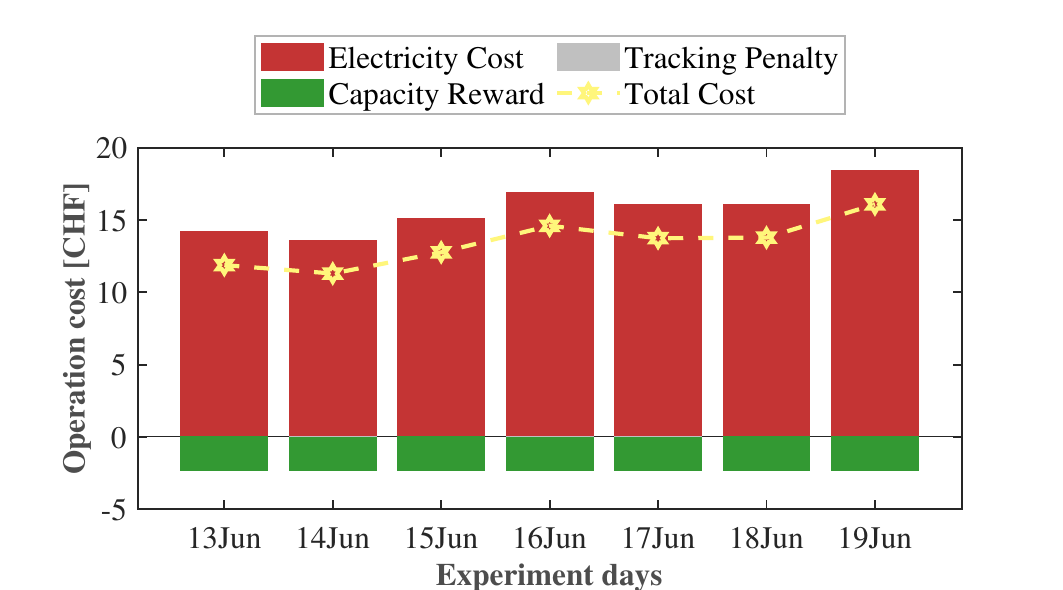}
    \caption{Different operating costs and the DR income of a continuous 7-day experiment: Case B - only the battery supports SFC  and the building operates by a default controller.}
    \label{fig:pdf_cost_bat}
\end{figure}

In this section, we compare the performance of the proposed SFC controller against two other scenarios using the default controller, under similar weather conditions. We evaluate three selected scenarios:
\begin{itemize}
    \item  \textbf{Scenario A:} This scenario involved 12 continuous days of operations under the proposed SFC controller, specifically from 30$^{th}$ July to 10$^{th}$ August 2022.
    \item  \textbf{Scenario B:} This scenario involved 7 continuous days (from 14$^{th}$ June to 20$^{th}$ June 2022). The building was controlled by the default controller, which maintained the indoor temperature around $24\degree C$. In addition, a simulated battery with the same specifications as in Scenario A provided the SFC service alone via the SFC controller, only considering the battery's dynamics.
    \item   \textbf{Scenario C:} This scenario also involved the same 7 continuous days as Scenario B, but without the battery, and the building did not provide SFC service.
\end{itemize}
To ensure a fair comparison, the weather conditions across all three scenarios were similar, as outlined in Table~\ref{tab:DR_compare}.
Figure~\ref{fig:pdf_cost_bat} shows the daily operating costs of Scenario B, with separate plots for different cost components.

\begin{figure}[!ht]
    \centering
    \includegraphics[width=1.0\linewidth]{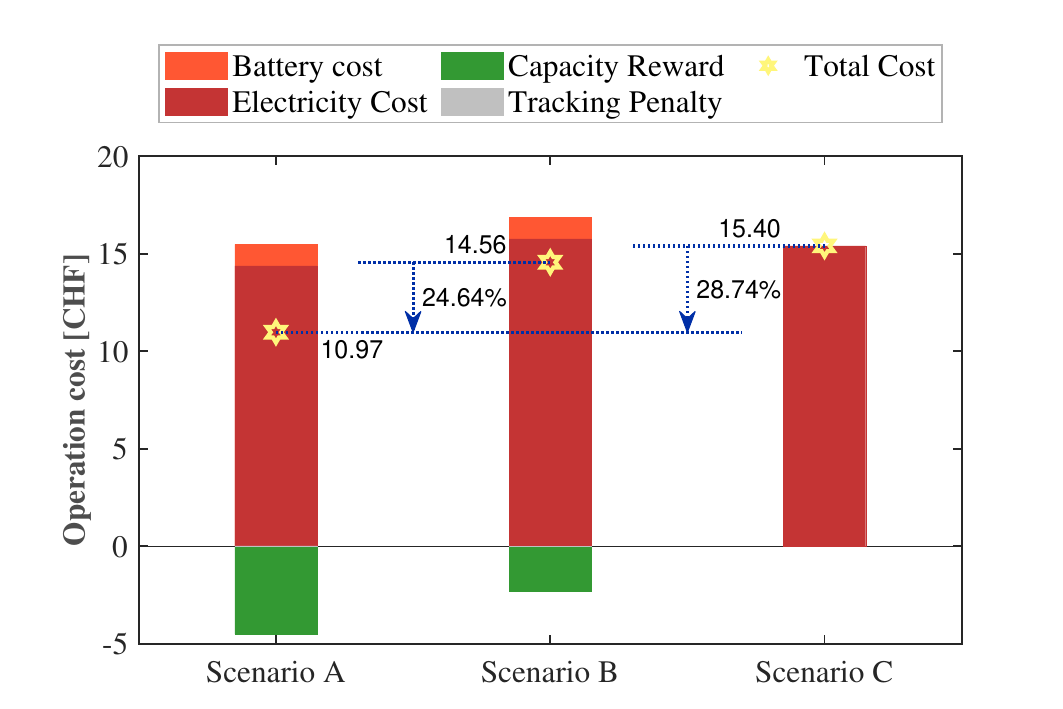}
    \caption{Comparison of average operating costs among Scenario A, B and C.}
    \label{fig:pdf_cost_compare}
\end{figure}

Figure~\ref{fig:pdf_cost_compare} compares the average operating costs among Scenarios A, B, and C. Both Scenarios A and B used battery system, so an estimated average battery cost was added, based on the market capacity price of a Tesla Powerwall 2 (813 CHF/kWh with a lifespan of 10 years). Scenario A, with a daily cost of 10.97 CHF, led to a 24.64\% and 28.74\% decrease compared to Scenarios B (14.56 CHF) and C (15.40 CHF), respectively.

\begin{figure}[!ht]
    \centering
    \includegraphics[width=1.0\linewidth]{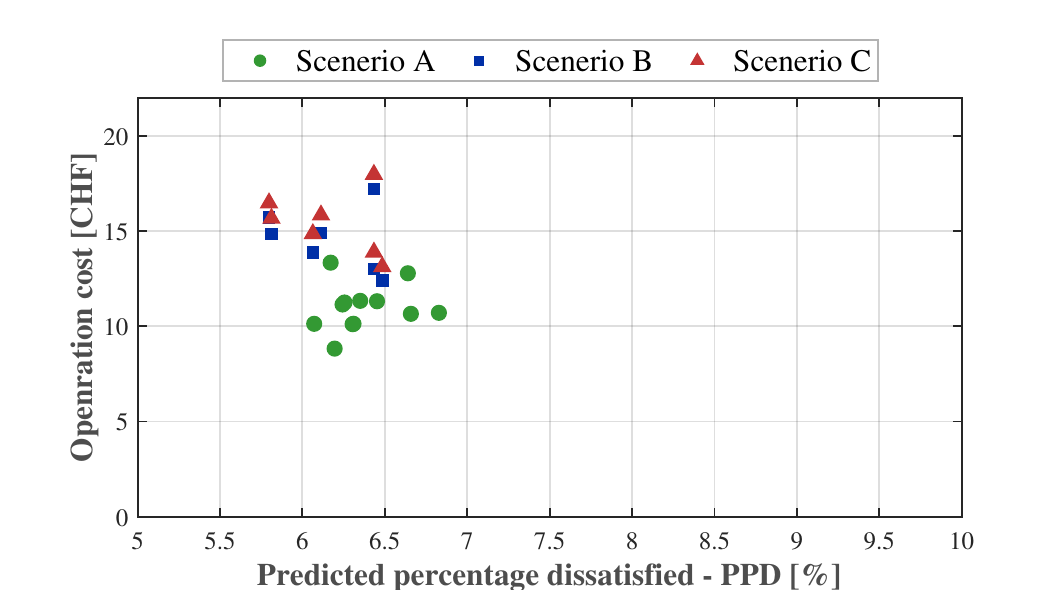}
    \caption{Comparison of average operating costs and comfort levels among Scenario A, B and C.}
    \label{fig:pdf_comfort_cost}
\end{figure}

We also compare the occupant comfort across the three scenarios using the \ac{ppd}, which estimates the proportion of a population's members who would be dissatisfied. Based on the Fanger comfort model~\cite{fanger1970thermal}, the \ac{pmv} was computed for each sampling period, choosing parameters of 0.1 m/s air velocity, 50\% indoor air relative humidity, 0.5 clo clothing resistance for summer, 1.2 met metabolic activity, and 0 met external work. Since the windows in the Polydome were always covered by curtains and the cooling air was evenly distributed via fabric pipes on the ceiling, we assumed the mean radiant temperature to be the same as the air temperature. The corresponding PPD [\%] was computed as a function of PMV:
\begin{align*}
    \text{PPD}(\text{PMV}) = 100-95e^{-0.03353\text{PMV}^{4}-0.2179\text{PMV}^{2}} \in \left[5, 100\right). 
\end{align*}
Figure~\ref{fig:pdf_comfort_cost} plots each day as a point based on its average PPD and operating cost, with different colors distinguishing the three scenarios. The PPDs for all scenarios were very similar and met the general comfort requirements specified by European standards~\cite{cen200715251}, i.e., $\text{PPD}<10\%$. For daily operating costs, most days in Scenario A were significantly lower than those in Scenarios B and C. 


\changeC{
In conclusion, the proposed SFC controller achieved substantial operational cost savings compared to standard control schemes while maintaining comfortable indoor conditions.
This result strongly confirms the efficiency and effectiveness of our proposed DDP and bi-level DeePC method.
}


\begingroup
\renewcommand{\arraystretch}{1.5} 

\begin{table}[!ht]
\centering
\footnotesize
\begin{tabular}{b{1.5cm}  b{0.7cm} b{0.7cm} b{0.7cm} b{0.7cm} b{0.7cm} b{0.7cm}
ccccccc}
\hline
   & \multicolumn{3}{c}{Indoor temperature $[^\circ C]$ } & \multicolumn{3}{c}{Solar radiation $[10W/m^2]$} \\ 
   & minimal & average & maximal & minimal & average & maximal \\
   \hline
\shortstack{Scenario A} & 18.31 & 24.77 & 33.69 &  0.00 & 26.82 & 87.68   \\
\shortstack{Scenario B\&C} & 17.10 & 24.63 & 32.88 &  0.00 & 26.59& 91.71 \\\hline
\end{tabular}
\caption{ Weather statistics}
\label{tab:DR_compare}
\end{table}

\endgroup

\section{Discussion} \label{sect:VI}

\subsection{Comparison with two other building controllers}
In the Polydome, two other predictive controllers have been implemented. In August 2017, a similar SFC controller using a state-space model provided the SFC service for 9 days~\cite{fabietti2018multi}. Compared to the default controller operating for 23 days with an ESS providing SFC service alone, the proposed method led to a 26\% reduction in operational costs.
In June and July 2021, a bi-level DeePC managed the building climate for 20 consecutive days~\cite{lian2021adaptive}. This resulted in an 18.4\% reduction in electrical power consumption compared to the default controller under similar weather conditions,  excluding ventilation power. However, due to varying outdoor weather conditions over the three years, direct cost comparisons are not feasible. Indeed, the average temperatures in Lausanne during the three experimental periods were 20.5$\degree C$, 23.1$\degree C$, and 24.8$\degree C$, respectively. 

Regarding the prediction of building indoor climate, the controllers used in 2021 and in this study employed the bi-level DeePC, while the 2017 controller relied on a parametric state-space model. 
The 2017 experiment used three-week data to identify the state-space model. In contrast, the DDP required fewer data (2 days of operational data in 2021; 5 days for the predictive building controller and 10 days for the day-ahead planner in this study). The consistency in parameter tuning also supports human-free adaptive updates, as shown in Section~\ref{sect:III_tune}.
Moreover, the DDP directly computes the input based on recent outputs, while the state-space model requires an additional Kalman Filter to identify the current state. Notably, compared to the 2021 experiment, this study conducted a more complicated and longer-term DR case study, further validating the efficacy and flexibility of the bi-level DeePC. 




\subsection{Future work}
Several potential extensions of this work are worth exploring. Firstly, the adaptive updating capability of DeePC can be further refined. While Section~\ref{sect:III_tune} recommends a larger $T$ for improved prediction, additional analyses could estimate an upper bound for $T$ or incorporate a forgetting factor for older data~\cite{shi2024efficient}, considering the slow time variability of building dynamics. It would be valuable to apply an adaptive bi-level DeePC across continuous seasons with automatic Hankel matrix updates. Indeed, seasonal building control experiments are rare in the published research, with only a few examples such as~\cite {sturzenegger2015model,granderson2018field,zhang2022model,banjac2023implementation}.

Secondly, incorporating physical rules in the bi-level DeePC method could enhance reliability and control performance. There has been growing interest in the building control community in combining physical knowledge with black-box models~\cite{di2022physically,bunning2022physics,lian2023physically}. In Section~\ref{sect:III_ada}, the Hankel matrix is updated only if a relaxed physical rule is satisfied. It is verified in~\cite{lian2023physically} that ignoring such a rule can lead to an overestimation of the flexibility bidding in the day-ahead planner.
In~\cite{lian2023physically}, we also proposed a physically consistent filter for the bi-level DeePC, which demonstrated superior prediction performance compared to an autoregressive exogenous model subject to the same physical rule when tested on operational data from the Polydome. Therefore, evaluating the closed-loop control performance in building simulations and real-world experiments could be highly promising.


\changeD{
Thirdly, the DeePC-based SFC controller used in this building-level case study could be extended to building clusters with heterogeneous HVAC equipment. DeePC offers feasible solutions for two key challenges in this extension: modeling and computational complexity. DeePC has demonstrated strong prediction and control performance across various types of buildings, HVAC devices and storage systems~\cite{chinde2022data,yin2024data,lian2021adaptive}. For example, \cite{yin2024data} validated that their DeePC variant was effective in controlling building climate with radiators, domestic hot water heating, and a stationary electric battery. The adaptive update for DeePC in this study also showed potential for further improving modeling capabilities, while the bi-level structure reduced tuning efforts. Although computational complexity increases as more systems are involved, decoupled dynamics and the linearity of DeePC enable efficient distributed algorithms~\cite{wang2023distributed}. 
}


\section{Conclusion} \label{sect:VII}
This paper presents the DDP and bi-level DeePC for various control targets. This data-driven scheme adapts to the latest operational data and requires less tuning effort compared to the standard DeePCs. The SFC case study, a complex DR scenario, highlighted DDP's efficiency and effectiveness in commissioning. We utilized the DDP and bi-level DeePC to design a hierarchical SFC controller, validated through a 52-day experiment at the Polydome testbed. This controller managed the combined system's flexibility for daily SFC operations, achieving a significant reduction in operating costs—24.64\% and 28.74\% compared to other control schemes.

\section*{CRediT authorship contribution statement}
\textbf{Jicheng Shi:} Conceptualization, Methodology, Software, Formal analysis, Investigation, Data curation, Writing - original draft, Visualization.
\textbf{Yingzhao Lian:} Conceptualization, Methodology, Software, Validation, Investigation, Writing - Review \& Editing.
\textbf{Christophe Salzmann:} Software, Resources, Writing - Review \& Editing, Project administration.
\textbf{Colin N. Jones:} Conceptualization, Methodology, Validation, Writing - Review \& Editing, Supervision, Funding acquisition.



\section*{Declaration of Generative AI and AI-assisted technologies in the writing process}
During the preparation of this work, the authors used ChatGPT in order to improve the readability and language and to check spelling and grammar. After using this tool, the authors reviewed and edited the content as needed and take full responsibility for the content of the publication.

\section*{Acknowledgement}
This work was supported by the Swiss National Science Foundation (SNSF) under the NCCR Automation project, grant agreement 51NF40 180545.



 \bibliographystyle{elsarticle-num} 
 \bibliography{ref.bib}





\end{document}